\documentclass[aps,letterpaper,floatfix,twocolumn]{revtex4-1}
\usepackage{graphicx,amssymb,amsmath}

\begin{document}

\title{Noise control in gene regulatory networks with negative feedback}

\author{Michael Hinczewski}
\affiliation{Department of Physics, Case Western Reserve University, OH 44106}

\author{D. Thirumalai}
\affiliation{Department of Chemistry, The University of Texas at Austin, TX 78712}

\begin{abstract}
  Genes and proteins regulate cellular functions through complex
  circuits of biochemical reactions.  Fluctuations in the components
  of these regulatory networks result in noise that invariably
  corrupts the signal, possibly compromising function.  Here, we
  create a practical formalism based on ideas introduced by Wiener and
  Kolmogorov (WK) for filtering noise in engineered communications
  systems to quantitatively assess the extent to which noise can be
  controlled in biological processes involving negative feedback.
  Application of the theory, which reproduces the previously proven
  scaling of the lower bound for noise suppression in terms of the
  number of signaling events, shows that a tetracycline
  repressor-based negative-regulatory gene circuit behaves as a WK
  filter.  For the class of Hill-like nonlinear regulatory functions,
  this type of filter provides the optimal reduction in noise.  Our
  theoretical approach can be readily combined with experimental
  measurements of response functions in a wide variety of genetic
  circuits, to elucidate the general principles by which biological
  networks minimize noise.
\end{abstract}

\maketitle

The genetic regulatory circuits that control all aspects of life are
inherently stochastic.  They depend on fluctuating populations of
biomolecules interacting across the crowded, thermally agitated
interior of the cell.  Noise is also exacerbated by low copy numbers of
particular proteins and mRNAs, as well as variability in the local
environment~\cite{Becskei00,Thattai01,Swain02,Ozbudak02,Elowitz02,Paulsson04}.
Yet the robust and reproducible functioning of key systems requires
mechanisms to filter out fluctuations. For example, regulating noise
is relevant in stabilizing cell-fate decisions in embryonic
development~\cite{Arias06}, prevention of random switching to
proliferating states in cancer-regulating miRNA
networks~\cite{Tsang07}, and maximization of the efficiency of
bacterial chemotaxis along attractant gradients~\cite{Andrews06}.
Comprehensive analysis of yeast protein expression reveals that
proteins involved in translation initiation, ribosome formation, and
protein degradation, have lower relative noise levels~\cite{Newman06},
suggesting natural selection could favor noise reduction for certain
essential cellular components~\cite{Fraser04,Lehner08}.

A common regulatory motif capable of suppressing noise is the negative
feedback
loop~\cite{Becskei00,Thattai01,Simpson03,Austin06,Dublanche06,Cox06,Zhang09,Singh09},
as has been explicitly demonstrated in synthetic gene
circuits~\cite{Becskei00,Austin06,Dublanche06}.  Feedback pathways for
a given chemical species can be mediated by numerous signaling
molecules, each with its own web of interactions and stochastic
characteristics that determine the ultimate effectiveness of the
system in damping the fluctuations of the target population and
maintaining homeostasis.  Thus, uncovering generic laws governing the
behavior of such control networks is difficult.  A major advance was
made by Lestas, Vinnicombe, and Paulsson 
(LVP)~\cite{Lestas10}, who showed that information theory can set a
rigorous lower bound on the magnitude of fluctuations within an
arbitrarily complicated homeostatic negative feedback network.  Since
the bound scales like the fourth root of the number of signaling
events, noise reduction is extremely expensive.  This underscores the
pervasiveness of biological noise, even in cases where there may be
evolutionary pressure to minimize it.

The existence of a rigorous bound raises a number of intriguing
issues.  Can a biochemical network actually reach this lower bound,
and thus optimally suppress fluctuations?  What would be the dynamic
behavior of such an optimal system, and how would it depend on the
noise spectrum of the system components?  Here we answer these
equations using a theory related to the optimal linear noise-reduction
filter, developed by Wiener~\cite{Wiener49} and
Kolmogorov~\cite{Kolmogorov41}.  Though the original context of
Wiener-Kolmogorov (WK) filter theory was removing noise from corrupted
signals in engineered communications systems, it has recently become a
powerful tool for characterizing the constraints on signaling in
biochemical networks~\cite{Hinczewski2014,Becker2015}.  Recently, we
showed that the action of kinase and phosphatase enzymes on their
protein substrates, the basic elements of many cellular signaling
pathways, can in fact effectively be represented as an optimal WK
filter~\cite{Hinczewski2014}.  The WK theory also describes how
systems like {\it E. coli} chemotaxis can optimally anticipate future
changes in concentrations of extracellular ligands~\cite{Becker2015}.
Although the classic WK theory is strictly defined for linear
filtering of continuous signals (a reasonable approximation for
certain biochemical networks), it can also be extended to yield
constraints in the more general case of nonlinear production of
molecular species with discrete population
values~\cite{Hinczewski2014}.

Interestingly, for a broad class of systems the WK linear solution
turns out to be the global optimum among all nonlinear or linear
networks, allowing us to delineate where nonlinearity is potentially
advantageous in biochemical noise control. Most importantly, since the
WK theory is formulated in terms of experimentally accessible dynamic
response functions, it also provides a design template for realizing
optimality in synthetic circuits.  As an illustrative example, we
predict that a synthetic autoregulatory TetR loop, engineered in
yeast~\cite{Nevozhay09}, can be fine-tuned to approximate an optimal
WK filter for TetR mRNA levels.  Though a simple design, similar
filters could be employed in nature to cope with Poisson noise arising
from small copy numbers of mRNAs, often on the order of 10 per
cell~\cite{Schwanhausser11}. Based on the application of the theory to
the synthetic gene network we propose that the extent of noise
reduction in biological circuits is determined by competing factors
such as functional efficiency, adaptation, and robustness.

\section*{Results}

To make the paper readable and as self-contained as possible many of
the details of the calculation are relegated to four Appendices.  The
main text contains only the necessary details needed to follow the
results without the distraction of the mathematics.

\subsection*{Linear response theory for a general control network}

To motivate the WK approach for a general control network, we start
with the simple case where two species within the network are
explicitly singled out~\cite{Lestas10}: a target $R$ with time-varying
population $r(t)$ fluctuating around mean $\bar{r}$, and one of the
mediators in the feedback signaling pathway $P$, with population
$p(t)$ varying around $\bar{p}$. We assume a continuum Langevin
description of the
dynamics~\cite{Gillespie00,Simpson03,Cox06,DeRonde10}, where the rate
\begin{equation}\label{eq1}
\dot\alpha(t) = k_\alpha(t) + n_\alpha(t)
\end{equation}
for $\alpha = r$ or $p$, can be broken down into deterministic
($k_\alpha$) and stochastic ($n_\alpha$) parts.  The function
$k_\alpha(t)$ encapsulates the entire web of biochemical reactions
underlying synthesis and degradation of species $\alpha$, and can be
an arbitrary functional of the past history of the system up to time
$t$.  It is typically divided into two parts, $k_\alpha(t) =
k^{+}_\alpha(t) - k^{-}_\alpha(t)$, corresponding to the production (+)
and destruction (-) rates of the species $\alpha$.  The term
$n_\alpha(t)$ is the additive noise contribution, which can also be
divided into two parts, $n_\alpha(t) = n_\alpha^\text{int}(t) +
n_\alpha^\text{ext}(t)$.  The first is the ``intrinsic'' or shot
noise, arising from the stochastic Poisson nature of $\alpha$
generation, $n^\text{int}_\alpha(t) = \sqrt{2 \bar{k}_\alpha}
\eta_\alpha(t)$, where $\bar{k}_\alpha$ is the mean production rate,
or equivalently the mean destruction rate, $\bar{k}_\alpha =
\overline{k^{+}_\alpha(t)} = \overline{k^{-}_\alpha(t)}$, and
$\eta_\alpha(t)$ is a Gaussian white noise function with correlation
$\overline {\eta_\alpha(t) \eta_{\alpha^\prime}(t^\prime)} =
\delta_{\alpha\alpha^\prime} \delta(t-t^\prime)$.  The second part,
$n^\text{ext}_\alpha(t)$, is ``extrinsic'' noise, which arises due to
fluctuations in cellular components affecting the dynamics of $R$ and
$P$ that are not explicitly taken into account in the two-species
picture.  These could include mediators in the signaling pathway, or
global factors like ribosome and RNA polymerase levels.  For
simplicity, our main focus will be the case of no extrinsic noise.
However, we will show later how a straightforward extension of the
theory reveals that the same system can behave like an optimal WK
filter under a variety of extrinsic noise conditions.

\begin{figure}[t]
\centerline{\includegraphics[width=\columnwidth]{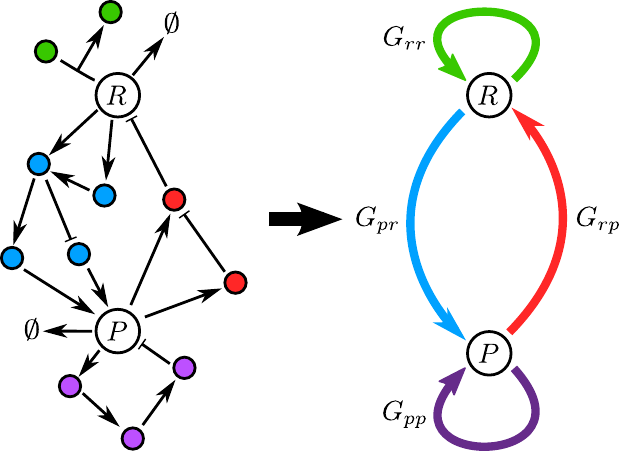}}
\caption{Schematic of a complex signaling network with
  the target species $R$ and one mediator $P$ singled out.  In
  focusing on two species, the action of all the other components is
  effectively encoded in four response functions---$G_{rr}(t)$,
  $G_{pp}(t)$, $G_{rp}(t)$, $G_{pr}(t)$---that describe how the entire
  dynamical system responds to perturbations in $R$ and $P$.}\label{net}
\end{figure}

For small deviations $\delta \alpha(t) = \alpha(t) - \bar{\alpha}$
from the mean populations $\bar{\alpha}$, $k_\alpha(t)$ can be
linearized with respect to $\delta \alpha(t)$,
\begin{equation}\label{eq2}
k_\alpha(t) = \sum_{\alpha^\prime = r,p} \int_{-\infty}^t
dt^\prime\,G_{\alpha \alpha^\prime}(t-t^\prime) \delta
\alpha^\prime (t^\prime),
\end{equation}
where $G_{\alpha \alpha^\prime}(t)$ are linear response functions,
which express the dependence of $k_\alpha(t)$ on the past history of
$\delta \alpha^\prime(t)$.  The functions
$G_{\alpha \alpha^\prime}(t)$ capture the essential characteristic
responses of the control network to perturbations away from
equilibrium (Fig.~\ref{net}).  In the static limit,
$G_{\alpha \alpha^\prime}(t)$ have appeared in various guises as
gains~\cite{Paulsson04}, susceptibilities~\cite{Zhang09}, or
steady-state Jacobian matrices~\cite{DeRonde10}, and in the
frequency-domain as loop transfer functions~\cite{Simpson03,Cox06}.
Feedback between $R$ and $P$ is encoded in the cross-responses $G_{rp}(t)$
and $G_{pr}(t)$.  In the simplest scenario, the only non-zero
self-responses $G_{\alpha\alpha}(t)$ are decay terms,
$G_{\alpha\alpha}(t) = -\tau^{-1}_\alpha \delta(t)$, where
$\tau_\alpha$ is the decay time scale for species $\alpha$.  However,
the theory works generally for more complicated self-response
mechanisms.

\subsection*{Control network as a noise filter} 

The connection between the linearized dynamical description and WK
filter theory arises from comparing the original system to the case
where feedback is turned off (i.e.~setting $G_{rp}(\omega)$ or
$G_{pr}(\omega)$ to zero).  Let us define a few terms to make the
noise filter analogy clear.  Without feedback, the target fluctuations
are $\delta r_0(t) \equiv s(t)$, where we denote $s(t)$ the {\it
  signal}.  This is to distinguish it from $\delta r(t)$ in the
original system, which is the {\it output}.  The difference between
the two, which reflects the impact of the feedback network, we
express as $\delta r(t) = s(t) - \tilde{s}(t)$, where $\tilde{s}(t)$
is referred to as the {\it estimate}.  In this analogy, minimizing
$\delta r(t)$ requires a feedback loop where the estimate
$\tilde{s}(t)$ is as close as possible to the signal $s(t)$.  The only
thing left to specify is the relationship between $\tilde{s}(t)$ and
$s(t)$.

The dynamical system in Eqs.~\eqref{eq1}-\eqref{eq2} takes a simple
form in Fourier space, where the fluctuations $\delta \alpha(\omega)$
satisfy:
\begin{equation}\label{eq3}
-i \omega \delta \alpha(\omega) = \sum_{\alpha^\prime = r,p} G_{\alpha
  \alpha^\prime}(\omega) \delta \alpha^\prime(\omega) +
n_\alpha(\omega).
\end{equation}
We solve Eq.~\eqref{eq3} for $\delta r(\omega)$ and break up the $R$
fluctuation into two contributions, $\delta r(\omega) = s(\omega) -
\tilde{s}(\omega)$, with the signal $s(\omega)$ and 
  estimate $\tilde{s}(\omega)$ given by:
\begin{equation}\label{eq4}
s(\omega) = -\frac{n_r(\omega)}{G_{rr}(\omega)+i \omega}, \quad \tilde{s}(\omega) = H(\omega)\left[s(\omega) + n(\omega)\right].
\end{equation}
Here we have introduced a noise function $n(\omega)$,
\begin{equation}\label{eq4b}
n(\omega) = \frac{n_p(\omega)}{G_{pr}(\omega)},
\end{equation}
and a filter function $H(\omega)$:
\begin{equation}\label{eq5}
 H(\omega) \equiv
\frac{G_{rp}(\omega)G_{pr}(\omega)}{G_{rp}(\omega)G_{pr}(\omega) -
  (G_{rr}(\omega)+i \omega)(G_{pp}(\omega)+i \omega)}.
\end{equation}
Thus in the time domain the estimate $\tilde{s}(t)$ is the convolution
of the filter function $H(t)$ and a noise-corrupted signal
$y(t) \equiv s(t) + n(t)$,
\begin{equation}\label{eq5ex}
\tilde{s}(t) = \int_{-\infty}^\infty dt^\prime H(t-t^\prime)
y(t^\prime ).
\end{equation}
Eqs.~\eqref{eq4}-\eqref{eq5} constitute a one-to-one mapping between
the linear response and noise filter descriptions of the system in
Fourier space.  They relate the four filter quantities, $s(\omega)$,
$\tilde{s}(\omega)$, $n(\omega)$, and $H(\omega)$, to the four linear
response functions $G_{rr}(\omega)$, $G_{rp}(\omega)$,
$G_{pr}(\omega)$, and $G_{pp}(\omega)$.

\begin{figure*}[t]
\centerline{\includegraphics[width=0.6\textwidth]{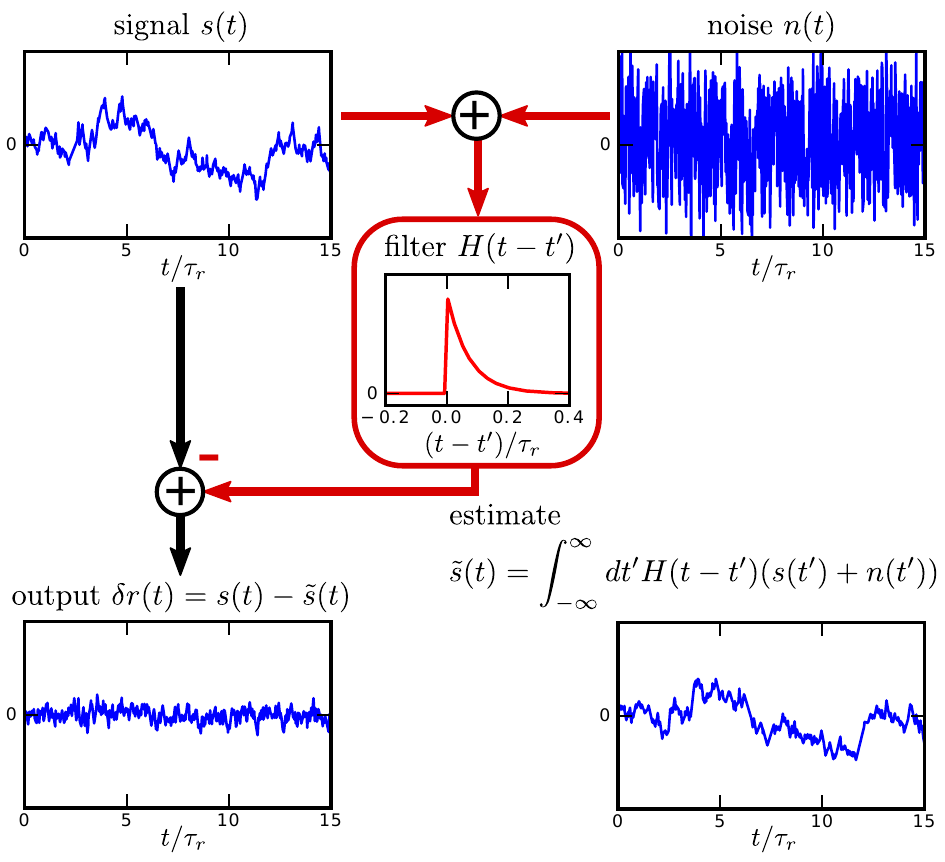}}
\caption{Signal processing diagram illustrating noise suppression in a
  negative feedback loop re-interpreted as a linear filter. The
  fluctuations in the target species $\delta r(t)$ (lower left) are
  expressed as $\delta r(t) = s(t) - \tilde{s}(t)$, where the raw
  signal $s(t)$ (upper left) equals $\delta r(t)$ in the absence of
  feedback control, and the estimate $\tilde{s}(t)$ (lower right) is
  the contribution of the feedback loop. This estimate is given by the
  convolution of a filter function $H(t)$ (center) and the corrupted
  signal $s(t) + n(t)$, where $n(t)$ is the noise (upper right).  The
   goal of Wiener-Kolmogorov theory is to find a causal
  $H(t)$ such that the standard deviation of $\delta r(t)$ is
  minimized.  All sample trajectories shown in the figure are
  generated from numerically solving the linearized version of
  the dynamical system in Eq.~\eqref{eq12}.}\label{fil}
\end{figure*}

The entire noise filter system is illustrated schematically in
Fig.~\ref{fil}.  Note that the noise function in the filter analogy,
$n(t)$, is related to $n_p(t)$ in Fourier space as
$n(\omega) = n_p(\omega)/G_{pr}(\omega)$.  Thus, the stochastic nature
of the mediator $P$ production makes estimation non-trivial,
since the function $H(t)$ must try to filter out the $n(t)$ component
in $y(t)$ in order to produce $\tilde{s}(t)$ close to $s(t)$.  Though
we confine ourselves throughout this work to the case of a dynamical
system with a single target and mediator species, one can easily
generalize the entire approach to explicitly include many mediators,
which could potentially be involved in a complex signaling pathway.
The linearized dynamical system in Eqs.~\eqref{eq1}-\eqref{eq2} would
still have the same form (with index $\alpha$ running over all the
species of interest), and the mapping onto the filter problem for the
target species would be analogous.  The only difference is that
$n(\omega)$ and $H(\omega)$ would be more complicated functions of the
various individual noise terms $n_\alpha(\omega)$ and the response
functions $G_{\alpha\alpha^\prime}(\omega)$ of the mediators.  In our
reduced, two species description, the action of all the unspecified
chemical components is effectively included in the four response
functions described above, with their stochastic effects contributing
to the extrinsic noise.  Fig.~\ref{net} shows a schematic of such a
reduction.  The fine-grained details of the signaling pathways
connecting our target $R$ and mediator $P$, potentially involving many
interacting species, are encoded in $G_{rr}$, $G_{pp}$, $G_{rp}$, and
$G_{pr}$.  As an example of how this two-species reduction would work
in practice, in Appendix B we treat an important
example of a feedback loop involving multiple mediators, representing
a signaling cascade in series.

\subsection*{Wiener-Kolmogorov theory yields the optimal filter}

The WK optimization problem consists of minimizing
$\sigma^2_r = \overline{(\delta r)^2}$, the variance of target
fluctuations, which are related to $H(t)$, $s(t)$, and $n(t)$ through
the frequency-domain integral~\cite{Bode50} (see derivation in
Appendix A):
\begin{equation}\label{eq6}
\sigma^2_r = \int_{-\infty}^{\infty} \frac{d\omega}{2\pi}\,
\left[|H(\omega)|^2 P_n(\omega) + |H(\omega)-1|^2 P_s(\omega) \right],
\end{equation}
where $H(\omega)$ is the Fourier transform of $H(t)$, and
$P_n(\omega)$, $P_s(\omega)$ are the power spectral densities (PSD) of
$n(t)$ and $s(t)$ respectively, i.e. the Fourier transforms of their
autocorrelation functions.  If $P_n(\omega)$ and $P_s(\omega)$ are
given, the task is to minimize $\sigma^2_r$ in Eq.~\eqref{eq6} over
all possible $H(\omega)$.  The main constraint that makes the solution
mathematically difficult is that $H(\omega)$ must correspond to a
physically realizable control network, which imposes the crucial
restriction that the time-domain convolution function $H(t)$ must be
causal, depending only on the past history of the input, $H(t) = 0$
for $t<0$.  The great achievement of Wiener and Kolmogorov was to
derive the form of the optimal causal solution $H_{\rm opt}(\omega)$:
\begin{equation}\label{eq7}
H_{\rm opt}(\omega) = \frac{1}{P_y^c(\omega)}\left\{ \frac{P_s(\omega)}{P_y^c(\omega)^\ast}\right\}_c.
\end{equation}
The $c$ super/subscripts refer to two different decompositions in the
frequency domain which enforce causality: (i) Any physical PSD, in
this case $P_y(\omega)$ corresponding to the corrupted signal
$y(t) = s(t) + n(t)$, can be written as
$P_y(\omega) = | P_y^c(\omega)|^2$.  The factor $P_y^c(\omega)$, if
treated as a function over the complex $\omega$ plane, contains no
zeros and poles in the upper half-plane
(${\rm Im}\, \omega > 0$)~\cite{ChaikinLubensky}.  (ii) We also define
an additive decomposition denoted by $\{F(\omega)\}_c$ (see Appendix
A) for any function $F(\omega)$, which consists of all terms in the
partial fraction expansion of $F(\omega)$ with no poles in the upper
half-plane.  In Appendix A we provide in detail a new derivation of
Eq.~\eqref{eq7}, the heart of the WK theory.

\subsection*{Optimal noise control in a yeast gene circuit with feedback}

\begin{figure*}
\begin{center}
\centerline{\includegraphics[width=\textwidth]{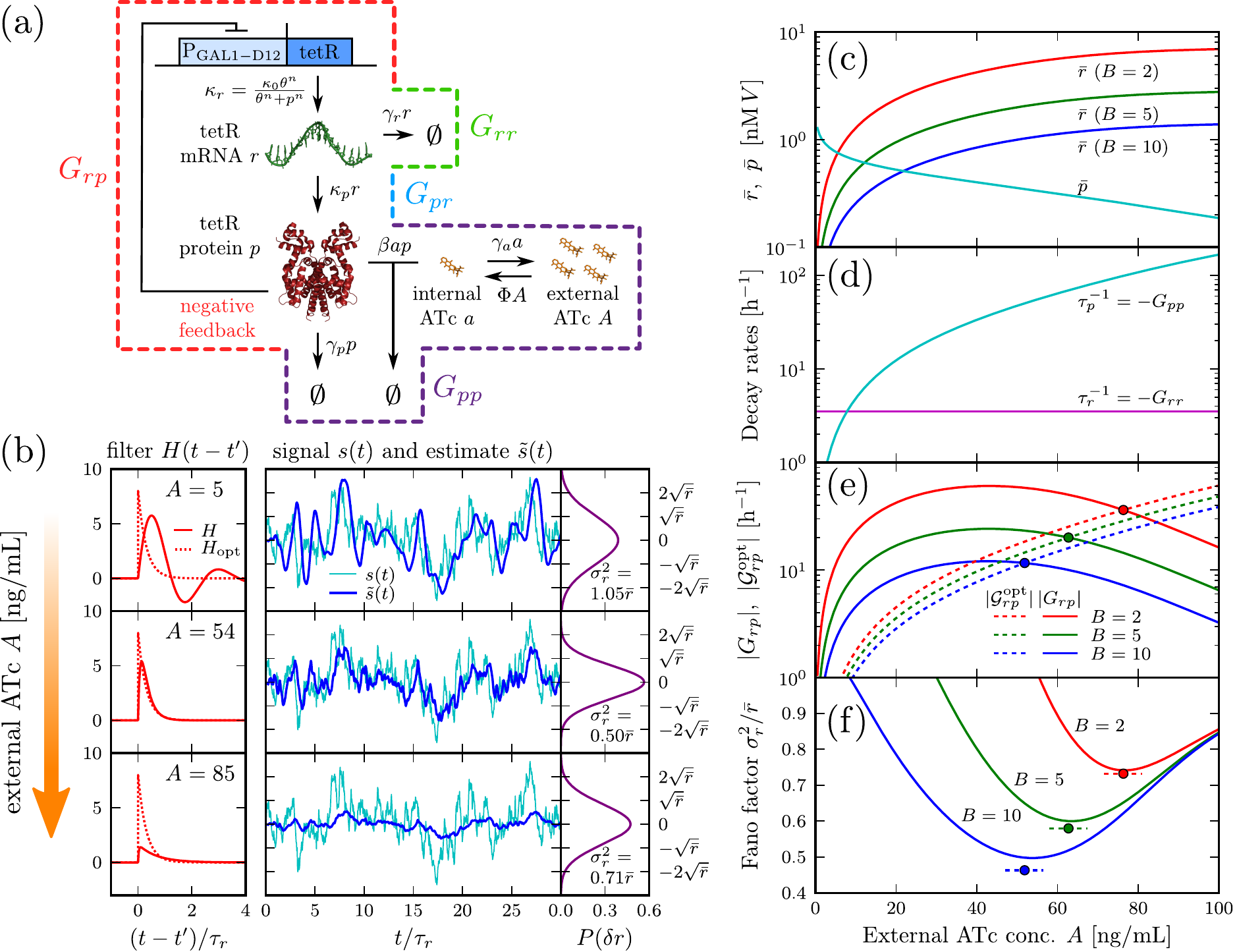}}
\caption{ (a) The synthetic yeast gene circuit designed by Nevozhay
  {\it et. al.}~\cite{Nevozhay09}.  The TetR protein negatively
  regulates itself by binding to its own promoter.  The inducer
  molecule ATc associates with TetR, inhibiting its repressor
  activity.  The subsequent panels show results for this gene circuit
  using the linear filter theory applied to the dynamical model of
  Eq.~\eqref{eq12}, with experimentally-derived parameters (Table~1).
  (b) Filter functions $H(t)$ and $H_\text{opt}(t)$, sample signal
  $s(t)$ and estimate $\tilde{s}(t)$ time series for burst ratio
  $B=10$ and three different values of extracellular ATc concentration
  $A$ [ng/mL].  $H(t)$ is from Eq.~\eqref{eq13b}, while
  $H_\text{opt}(t)$ is from Eq.~\eqref{eq10}.  The sample time series
  trajectories are numerical solutions of the linearized
  Eq.~\eqref{eq12}.  On the right are the resulting equilibrium
  probability distributions $P(\delta r)$, where $\delta r(t) = s(t) -
  \tilde{s}(t)$, which are Gaussians with variance $\sigma_r^2$.  For
  $A \approx 54$ ng/mL, the circuit approximately functions as an
  optimal WK filter ($H(t)$ is close to $H_\text{opt}(t)$), maximally
  suppressing fluctuations in the population levels of TetR mRNA
  (minimizing $\sigma_r^2/\bar{r}$).  (c) Mean populations of free
  intracellular TetR mRNA, $\bar{r}$, and TetR protein dimers,
  $\bar{p}$.  (d) The decay rates of free mRNA and proteins,
  $\tau_r^{-1}$ and $\tau_p^{-1}$, which are related to the network
  self-response functions $G_{rr}$ and $G_{pp}$ (both are constants in
  the frequency domain as shown in Eq.\ref{eq13}).  (e) The magnitude of the network
  cross-response, $|G_{rp}|$ (solid lines), plotted together with the
  optimal magnitude $|{\cal G}_{rp}^\text{opt}| = \tau_p^{-1}
  (1+\sqrt{1+B})^{-1}$ (dashed lines).  Filled circles mark the
  intersection defining $A = A_\text{opt}$, where the system behaves
  approximately like an optimal WK filter.  (f) The Fano factor
  $\sigma^2_r/\bar{r}$ (solid lines), compared to the optimal WK value
  $\sigma^2_{r,\text{opt}}/\bar{r} = 2/(1+\sqrt{1+B})$ (horizontal
  dashed lines).  Filled circles mark the position $A =
  A_\text{opt}$.}\label{tet}
\end{center}
\end{figure*}

To illustrate the nature of the optimal WK solution we choose as a
case study the yeast negative autoregulatory gene circuit designed by
Nevozhay {\it et. al.}~\cite{Nevozhay09}, drawn schematically in
Fig.~\ref{tet}(a). The gene encoding for the TetR protein is under the
control of the P$_{\rm GAL1-D12}$ promoter, whose activity can be
repressed by binding TetR dimers.  The strength of the feedback can be
modulated by changing the extracellular concentration $A$ of the
inducer anhydrotetracycline (ATc), which enters the cell, binds to
TetR and prevents its association with the promoter, thus weakening
repression.  

In order to analyze the TetR negative feedback gene circuit, we start
with the simple mathematical model introduced in
Ref.~\citenum{Nevozhay09}, which provided results that are consistent
with the experimental data. The simplified model, which captures the
essence of the synthetic gene network, features as the main variables
the population of free intracellular TetR dimer, $p(t)$, and free
intracellular ATc molecules, $a(t)$.  In addition to the regulatory
loop, the experimental gene circuit has a parallel yEGFP reporter
portion, which acts as a monitor of TetR protein levels.  Because we
focus on the system as a noise filter for the TetR mRNA population,
and the yEGFP part does not influence this analysis~\cite{Nevozhay09},
we ignore the reporter circuit.

The production of the TetR dimers occurs in a single step, with the
autoregulation of the rate described by a repressory Hill function.
We divide this step into two parts, introducing as an additional
variable the population of TetR mRNA $r(t)$.  The feedback loop
(Fig.~\ref{tet}(a)) consists of mRNA production at a rate
given by the Hill function $\kappa_r(t) = \kappa_0 \theta^n/(\theta^n
+ p^n(t))$, followed by TetR dimer generation at a rate given by
$\kappa_p r(t)$.  The degradation/dilution of the mRNA and dimers is
modeled through decay terms $\gamma_r r(t)$ and $\gamma_p p(t)$.  We
could have modeled additional (comparatively fast) chemical substeps
involved in this loop, such as TetR dimerization, the binding of the
repressor to the individual promoter sites, or the role of RNAP and
ribosomes in the transcription and translation processes.  Though we
limit ourselves to the two substep description to illustrate the
filter theory, the stochastic effects of additional complexity can be
approximately treated through general ``extrinsic'' noise terms
incorporated into $n_r(t)$ and $n_p(t)$.

The main experimental variable that allows tuning of the yeast gene network 
behavior is the external ATc concentration $A$, which is assumed to be
time independent.  As illustrated in Fig.~2(a), there is an
influx $\Phi A$ of ATc molecules into the cell.  Once inside, the ATc
molecules associate with the TetR at a rate $\beta a(t) p(t)$. Additional
loss of intracellular ATc through degradation, outflux, and
dilution is modeled through an effective decay rate $\gamma_a a(t)$.
We assume that the dissociation of ATc from TetR occurs on long enough timescales
 that it can be ignored.  Since the influx/association/outflux
of ATc is fast compared to the transcription and translation processes
of the main feedback loop, we further assume that $a(t)$
instantaneously equilibriates at the current value of $p(t)$.  Thus,
the dependence of $a(t)$ on $p(t)$ is determined by equating the
influx and total loss rate, which leads to $a(p(t)) = \Phi A /
(\gamma_a + \beta p(t))$.

For the model described above, the dynamical equations for $r(t)$ and
$p(t)$ are,
\begin{equation}\label{eq12}
\begin{split}
\dot{r}(t) &= -\gamma_r  r(t) + \frac{\kappa_0 \theta^n}{\theta^n + p^n(t)} + n_r(t),\\
\dot{p}(t) &= -\gamma_p  p(t) + \kappa_p r(t) - \frac{\beta \Phi A p(t)}{\gamma_a + \beta p(t)} + n_p(t).
\end{split}
\end{equation}
The parameters, with values derived from experimental
fitting~\cite{Nevozhay09}, are listed in Table~1.  The only quantity
that is not independently known from the fit is the rate $\kappa_p$,
which we allow to vary in the range
$\kappa_p/\gamma_r \equiv B = 2-10$, comparable to typical
experimentally measured protein burst sizes~\cite{Cai06}.  Setting the
right sides of Eq.~\eqref{eq12} to zero, and averaging over $n_r(t)$
and $n_p(t)$, we numerically solve for the equilibrium values
$\bar{r}$ and $\bar{p}$ as a function of external ATc concentration
$A$ [Fig.~\ref{tet}(c)].  For $A =0$, the promoter is nearly fully
repressed, but with increasing $A$, the mean population $\bar{p}$ of
free TetR dimers is reduced, weakening the repression and boosting the
mean mRNA population $\bar{r}$.  Changing $A$ allows us to explore a
wide range of control network behavior. Note that since $\bar{p}$
depends on $B$ only through the the product $\kappa_0 B$, and the
value of this product is fixed at a constant value from the experimental fit
(Table~\ref{par}), $\bar{p}$ is independent of $B$.  On the other
hand, $\bar{r}$, which is proportional to $\kappa_0$, is inversely
proportional to $B$.

\begin{table}[t]
\begin{minipage}[h]{\linewidth}\centering
\begin{tabular}{cc}
\hline
Parameter & Value\\
\hline
$n$ & 4\\
$\theta$ & $0.44\:{\rm nM}\,V$ \\
$\gamma_r$ & $3.5\:{\rm h}^{-1}$~\footnote{Ref.~\citenum{Garcia-Martinez04}}\\
$\gamma_p$ & $0.12\:{\rm h}^{-1}$ \\
$\gamma_a$ & $1.2\:{\rm h}^{-1}$\\
$\beta$ & $3.6\:{\rm nM}^{-1}{\rm h}^{-1}V^{-1}$\\
$\Phi$ & $0.6\:{\rm h}^{-1}V$\\
$\kappa_0$ & $50\:{\rm nM}\,\,{\rm h}^{-1}V B^{-1}$~\footnote{The burst ratio $B \equiv \kappa_p / \gamma_r$.
  Though not independently determined by the experimental fit, we
  assume that $B$ is in the range $B = 2-10$~\cite{Cai06}.}\\
$A$ & $0 - 500$ ng/mL~\footnote{For external ATc concentration $A$, 1 ng/mL corresponds to $2.25$ nM.}\\
\hline
\end{tabular}
\caption{Parameter values for the dynamical model of the yeast
  synthetic gene circuit (Eq.~\eqref{eq12}).  The cell volume $V$ is
  assumed fixed.  Unless otherwise noted, all values are taken from the
  experimental fit of Ref.~\citenum{Nevozhay09}.}\label{par}
\end{minipage}
\end{table}

Linearizing Eq.~\eqref{eq12} around $\bar{r}$ and $\bar{p}$, we
extract the following frequency-domain response functions:
\begin{equation}\label{eq13}
\begin{split}
G_{rr}(\omega) &= -\tau_r^{-1} = -\gamma_r, \quad G_{rp}(\omega) = -\frac{\kappa_0 n \theta^n \bar{p}^{n-1}}{(\theta^n + \bar{p}^n)^2},\\
G_{pp}(\omega) &= -\tau_p^{-1} = -\gamma_p - \frac{\beta \gamma_a \Phi A}{(\gamma_a+\beta \bar{p})^2}, \quad G_{pr}(\omega) = \kappa_p.
\end{split}
\end{equation}
All the functions are constants in the frequency domain.  Here
$\tau_r$ and $\tau_p$ are effective decay times for the mRNA and
proteins, respectively.  The value of $\tau_r$ is fixed, and sets the
intrinsic time scale of mRNA fluctuations, but $\tau_p$ and $G_{rp}$
depend on $\bar{p}$, which is a function of the external ATc
concentration $A$.  In fact, association with intracellular ATc,
described by the second term in the $G_{pp}$ expression above, is the
dominant form of decay for the free TetR dimers.  Fig.~\ref{tet}(d)
plots the effective decay constants $\tau_r^{-1}$ and $\tau_p^{-1}$ as
a function of $A$.  Except for $A \lesssim 8$ ng/mL we are in the
regime where $\tau_p^{-1} \gg \tau_r^{-1}$, which is relevant in
simplifying the optimality condition for $G_{rp}(\omega)$ discussed
below.

The optimal filter calculation for the TetR gene circuit depends on
the linear response functions of Eq.~\eqref{eq13}.  We obtain the
following power spectra for the signal and noise in the absence of
extrinsic noise:
\begin{equation}\label{si1}
P_s(\omega) = \frac{2 \bar{r} \tau_r}{1 + (\omega \tau_r)^2}, \quad P_n(\omega) = \frac{2\bar{r}\tau_r}{B},
\end{equation}
where the burst ratio $B \equiv \kappa_p \tau_r$ is the mean number of
proteins synthesized per mRNA during the lifetime $\tau_r$.  The
problem is to evaluate Eq.~\eqref{eq7} for $H_{\rm opt}(\omega)$.  The
sum of signal plus noise, $y(\omega) = s(\omega) + n(\omega)$, has a
power spectrum $P_y(\omega) = P_s(\omega) + P_n(\omega)$, which we can
rewrite as follows:
\begin{equation}\label{si2}
\begin{split}
P_y(\omega) &= 2 \bar{r} \tau_r \left[\frac{1}{1 + (\omega \tau_r)^2} + \frac{1}{B}\right]\\
&= \left|\left(\frac{2\bar{r}\tau_r}{B}\right)^{1/2} \frac{\sqrt{1+B} - i \omega \tau_r}{1- i\omega \tau_r }\right|^2.
\end{split}
\end{equation}
The expression within the absolute
value brackets is zero only at $\omega = -i \tau_r^{-1}
\sqrt{1+B}$, and has a simple  pole at $\omega = -i \tau_r^{-1}$.  Since
all the zeros and poles are in the lower complex $\omega$ half-plane,
it satisfies the criterion for the causal term in the factorization
$P_y(\omega) = |P_y^c(\omega)|^2$.  Thus:
\begin{equation}\label{si3}
P_y^c(\omega) = \left(\frac{2\bar{r}\tau_r}{B}\right)^{1/2} \frac{\sqrt{1+B} - i \omega \tau_r}{1- i\omega \tau_r }.
\end{equation}
The other causal term in Eq.~\eqref{eq7} involves the additive
decomposition $\left\{P_s(\omega)/P_y^c(\omega)^\ast\right\}_c$.  This is
calculated by looking at the partial fraction expansion of
$P_s(\omega)/P_y^c(\omega)^\ast$:
\begin{equation}\label{si4}
\begin{split}
\frac{P_s(\omega)}{P_y^c(\omega)^\ast} &= \frac{(2\bar{r} \tau_r B)^{1/2}}{(1-i \omega \tau_r) ( \sqrt{1+B} + i \omega \tau_r)}\\
&= \frac{(2\bar{r} \tau_r B)^{1/2}}{(1-i \omega \tau_r) ( \sqrt{1+B} + 1)}\\
&\quad + \frac{(2\bar{r} \tau_r B)^{1/2}}{(1+ \sqrt{1+B}) ( \sqrt{1+B} + i \omega \tau_r)}.
\end{split}
\end{equation}
Of the two terms in the partial fraction expansion, only the first 
has poles solely in the lower complex $\omega$ half-plane.
Hence, it is the only one that contributes to
$\left\{P_s(\omega)/P_y^c(\omega)^\ast\right\}_c$:
\begin{equation}\label{si5}
\left\{\frac{P_s(\omega)}{P_y^c(\omega)^\ast}\right\}_c = \frac{(2\bar{r} \tau_r B)^{1/2}}{(1-i \omega \tau_r) ( \sqrt{1+B} + 1)}.
\end{equation}
Inserting Eqs.~\eqref{si3} and \eqref{si4} into Eq.~\eqref{eq7}, we
finally find that the optimal filter is:
\begin{equation}\label{si6}
H_{\rm opt}(\omega) = \frac{\sqrt{1+B}-1}{\sqrt{1+B} - i \omega \tau_r}.
\end{equation}
Transforming $H_{\rm opt}(\omega)$ into the time domain, we find
\begin{equation}\label{eq10}
H_{\rm opt}(t) = \left(\tau_{\rm avg}^{-1} - \tau_r^{-1}\right) e^{-t/\tau_{\rm avg}} \Theta(t), 
\end{equation}
where $\tau_{\rm avg} = \tau_r/\sqrt{1+B}$, and $\Theta(t)$ is a unit
step function ensuring that the filter operates only on the past
history of its input.  For $B\gg 1$ the prefactor in Eq.~(\ref{eq10})
is $\approx \tau_{\rm avg}^{-1}$, and $H_{\rm opt}(t)$ has a
straightforward interpretation: it approximately acts as a moving
average of the corrupted signal $y(t) = s(t) +n(t)$ over a time scale
$\tau_{\rm avg}$.  In order to get the best estimate $\tilde{s}(t)$,
the averaging interval $\tau_{\rm avg}$ can neither be too long, since
it would blur out the features of the signal $s(t)$ (which vary on the
time scale $\tau_r$), nor too short, since it would be ineffective at
smoothing out the noise distortion $n(t)$.  Hence, there must exist an
optimum $\tau_{\rm avg}$, which is naturally proportional to $\tau_r$,
the main time scale for the mRNA.

In Fig.~\ref{tet}(b), we show how the noise filter properties of the
system vary with $A$ for a burst ratio of $B = 10$.  The filter
function $H(t)$ (solid red curve) differs substantially from
$H_\text{opt}(t)$ (dotted red curve) for large and small $A$, but
approaches the optimal form near $A = 54$ ng/mL.  Consequently, at
this value of $A$ we get the closest correspondence between the
plotted sample trajectories of signal $s(t)$ (cyan curve) and estimate
$\tilde{s}(t)$ (blue curve).  Similarly, the equilibrium probability
distribution of the output, $P(\delta r)$, shown to the right of the
trajectories, exhibits the smallest Fano factor $\sigma^2_r/\bar{r}$.
The latter is a measure of noise magnitude, and has a reference value
of unity if mRNA production was a pure Poisson process, as would be
the case without feedback.  Optimality is realized in the intermediate $A$
regime of partial repression, where the $R$ to $P$ responsiveness, as
measured by $|G_{rp}|$, is large.  Effective noise suppression
requires that $R$ be sensitive to changes in $P$, so that information
about $R$ fluctuations can be transmitted through the negative
feedback loop.  

In order to understand the optimality condition for $H(t)$ in more detail, let
us look at the explicit expression for $H(t)$ in the TetR
system, given by the inverse Fourier transform of Eq.~\eqref{eq5} with
the response functions of Eq.~\eqref{eq13}:
\begin{equation}\label{eq13b}
H(t) = \frac{G_{rp}\kappa_p}{\omega_1 -\omega_2}(e^{-\omega_1 t}- e^{-\omega_2 t})\Theta(t),
\end{equation}
where $\omega_1$, $\omega_2$ are the two $\omega$ roots of the
denominator in Eq.~\eqref{eq5}.  Assuming $\tau_p \ll \tau_r$ (which
holds good except for small values $A \lesssim 8$ ng/mL, as seen in
Fig.~\ref{tet}(d)), we can directly show the approach of $H(t)$ to
optimality at a specific intermediate value of $G_{rp}$.  When
$G_{rp}$ equals ${\cal G}^\text{opt}_{rp}(B,\tau_p) =
-1/(\tau_p(1+\sqrt{1+B}))$, the roots $\omega_1 \approx
\tau_\text{avg}^{-1}$, $\omega_2 \approx \tau_p^{-1} + \tau_r^{-1} -
\tau_\text{avg}^{-1}$, up to corrections of order $\tau_p/\tau_r^2$.
In this case, Eq.~\eqref{eq13b} becomes
\begin{equation}\label{eq13c}
H(t)|_{G_{rp}= {\cal G}^\text{opt}} \approx H_\text{opt}(t) \left[\frac{1-e^{-\left(\tau_p^{-1} +\tau_r^{-1}-2\tau_\text{avg}^{-1}\right)t}}{1+\tau_p(\tau_r^{-1}-2\tau_\text{avg}^{-1})} \right], 
\end{equation}
where the factor in the brackets on the right equals 1 in the limit
$\tau_p \to 0$ for all $t>0$.  Up to this correction factor, we thus
expect the system to behave optimally at $A = A_\text{opt}$, defined
by the condition $G_{rp} = {\cal G}^\text{opt}_{rp}(B,\tau_p)$, so
long as $A_\text{opt}$ is large enough to satisfy $\tau_p \ll \tau_r$.
Fig.~\ref{tet}(e) shows $G_{rp}$ and ${\cal G}^\text{opt}_{rp}$ curves
for $B = 2,\,5,\,10$, with dots marking the intersection points that
define $A_\text{opt}$ for each $B$.  As explained above, $|G_{rp}|$ is
small at small and large $A$, and reaches a maximum in between.  At
fixed $B$, $|{\cal G}^\text{opt}_{rp}(B,\tau_p)| \propto \tau_p^{-1}$,
so it increases monotonically with $A$, as larger concentrations of
the inducer increase the effective decay rate of free proteins.  Thus,
for each $B$ there is a single intersection point $A_\text{opt}$ at
an intermediate concentration of the inducer.

Fig.~\ref{tet}(f) shows the Fano factor $\sigma^2_r/\bar{r}$ versus
$A$ for various $B$.  As the control network approximates optimality
at $A_{\rm opt}$ for each $B$, the Fano factor nears its minimum,
close to the theoretical limit marked by the horizontal dashed lines.
This limit is the minimal possible $\sigma^2_r/\bar{r}$, calculated
from Eq.~(\ref{eq6}) using $H_{\rm opt}(t)$ from Eq.~(\ref{eq10}):
\begin{equation}\label{eq11}
\frac{\sigma^2_{r,{\rm opt}}}{\bar{r}} = \frac{2}{1+\sqrt{1+B}} \ge \frac{2}{1+\sqrt{1+4B}}
\end{equation}
A few comments concerning the above equation are in order.  (1) The
result on the far right-hand side is the rigorous lower bound derived
by LVP~\cite{Lestas10}. In their case, the feedback mechanism through
the rate function $k_r(t)$ could be any causal functional of $p(t)$,
linear or nonlinear.  The Fano factor of the optimal linear filter
differs in form only by the coefficient of $B$, and is always within a
factor of 2 of the lower bound for any value of $B$.  (2) For
Gaussian-distributed signal $s(t)$ and noise $n(t)$ time series, the
linear filter is optimal among all possible filters~\cite{Bode50}.  If
the system fluctuates around a single stable state, and the copy
numbers of the species are large enough that their Poisson
distributions converge to Gaussians (mean populations $\gtrsim 10$),
the signal and noise are usually approximately Gaussian.  This is a
wide class of systems where the rigorous lower bound (the last term in
Eq.~\ref{eq11}) can never be achieved.  In other words, here the WK
filter yields the most efficient feedback mechanism.  Although, as
pointed out by LVP, nonlinearity could lead to additional noise
reduction, the benefits are likely to be restricted to those systems
where the signal and/or noise are substantially non-Gaussian.
However, since the form of the optimal control network has not been
found in the general nonlinear case, it remains an interesting open
question whether the LVP bound can actually be reached even within
this category of systems.  We will return to this issue in the next
section.  (3) The parameter $B$ is the key determinant of noise
reduction.  For $B \ll 1$, there are not enough signaling events to
control the mRNA fluctuations, and as $B \to 0$ we approach
$\sigma^2_{r,{\rm opt}}/\bar{r} \to 1$, the no-feedback Poisson
result.  In the limit $B \gg 1$ signaling is effective, and the Fano
factor decreases with $B$ as
$\sigma^2_{r,\text{opt}}/\bar{r} \approx 2/\sqrt{B}$.  For large
enough $B$ we approach perfect control, but at extreme expense: the
standard deviation of the mRNA fluctuations
$\sigma_{r,\text{opt}} \propto B^{-1/4}$, the same scaling derived by
LVP.

\subsection*{WK theory constrains the performance of a broad class of nonlinear, discrete regulatory networks}

\begin{figure}[t]
\centering\includegraphics*[width=\columnwidth]{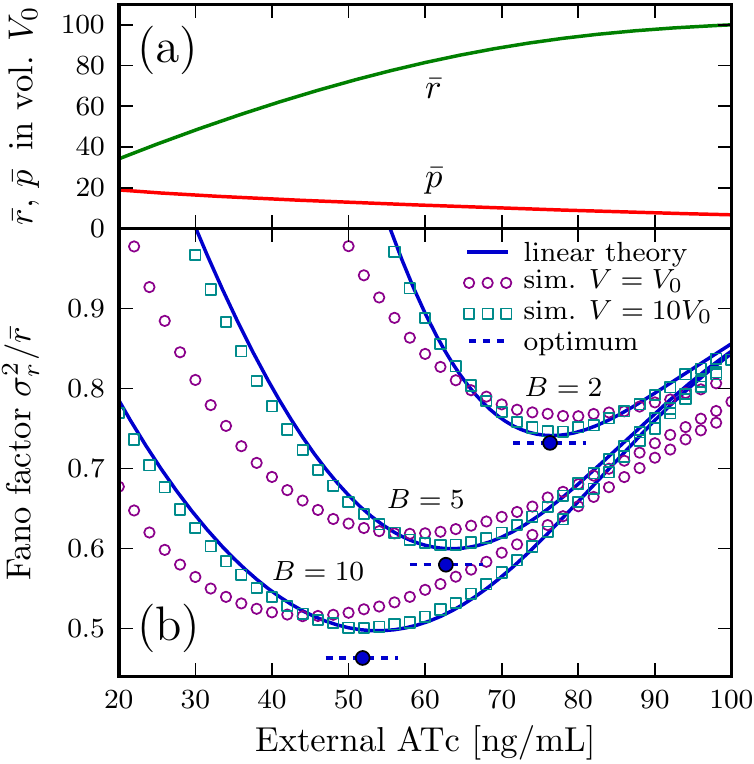}
\caption{Results of simulation and theory for the yeast synthetic gene
  circuit~\cite{Nevozhay09}, as a function of extracellular ATc
  concentration $A$. (a) Mean populations of free TetR mRNA $\bar{r}$
  and TetR dimer $\bar{p}$, assuming a cell volume $V_0 = 60$ fL.  (b)
  The Fano factor $\sigma^2_r/\bar{r}$ for burst factor $B = 2,5,10$,
  as predicted by the linear filter theory (solid lines), versus
  stochastic numerical simulations at two different volumes, $V = V_0$
  (circles) and $V = 10V_0$ (squares).  The WK filter theory predicts the
  minimal Fano factor $\sigma^2_{\rm opt}/\bar{r}$ given by
  Eq.~(\ref{eq11}) (horizontal dashed lines).  The system can be tuned
  to approach optimality near a particular $A_{\rm opt}$ obtained by
  the condition $G_{rp} = {\cal
    G}^{\rm opt}_{rp}$ (filled circles).}\label{sim}
\end{figure}

The results in Fig.~\ref{tet} rely on a linearized, continuum approach
to the TetR dynamical system. To assess if the conclusions based on
the WK optimal filter hold if these approximations are relaxed, we first
performed kinetic Monte Carlo simulations of the full nonlinear system
(Eq.~\eqref{eq12}) using the Gillespie algorithm~\cite{Gillespie77}.
We chose a cell volume of $V = V_0 = 60$ fL, within the observed range
for yeast~\cite{Jorgensen02}, which corresponds to the mean
populations $\bar{r}$ and $\bar{p}$ shown in Fig.~\ref{sim}(a) as a
function of $A$.  (For example, at $A = A_\text{opt} = 62.7$ ng/mL
when $B=5$, $\bar{r}\approx 84$ and $\bar{p} \approx 11$. In addition
to the nonlinearity, the discrete nature of the populations in the
simulation might play a role at these low copy numbers.)  The
numerical results for the Fano factor $\sigma^2_r/\bar{r}$ are plotted
in Fig.~\ref{sim}(b) at $B = 2,\,5,\,10$, for $V = V_0$ (circles) and
also for comparison at a larger volume $V = 10V_0$ (squares).  The
blue curves show the linear theory results, and the dashed lines are
the optimality predictions for $\sigma^2_{r,\text{opt}}/\bar{r}$.
Although nonlinearity and discreteness effects do change the results,
the linear theory gives a reasonable approximation, and the minimum is
still near $A_\text{opt}$.  The feedback mechanism is nonlinear in the
simulations, but it does not do better than the linear predictions for
$\sigma^2_{r,\text{opt}}/\bar{r}$ for the parameters used to describe
the experimental results.  Though the intrinsic population noise is
Poisson-distributed in the simulations, the Poisson distribution is
very close to Gaussian, even for copy numbers as low as
$\sim {\cal O}(10)$.  Since the linear filter is the true optimum for
a Gaussian-distributed signal and noise~\cite{Bode50}, we do not
expect improvements in noise suppression by employing a nonlinear
version.  In the opposite limit of large copy numbers, $V \to \infty$,
the continuum approximation should be valid, and population
fluctuations increasingly negligible relative to the mean.  Thus, the
linear theory should directly apply in this limit, and indeed we see
that for $V = 10V_0$ the discrepancies between numerical and theory
results are substantially reduced (Fig.~\ref{sim}(b)).  It is worth
emphasizing, that even at the realistically small cell volume $V_0$,
the linear theory retains much of its predictive power.  More
generally, the conditions for WK optimality do not have to be
perfectly satisfied in order for the filter to perform close to
maximum efficiency.  There is an inherent adaptability and robustness
in near-optimal networks, as reflected in the broad minima of
$\sigma^2_r/\bar{r}$ as a function of $A$ (Fig.~\ref{sim}(b)).

\begin{figure}[t]
\centering\includegraphics[width=\columnwidth]{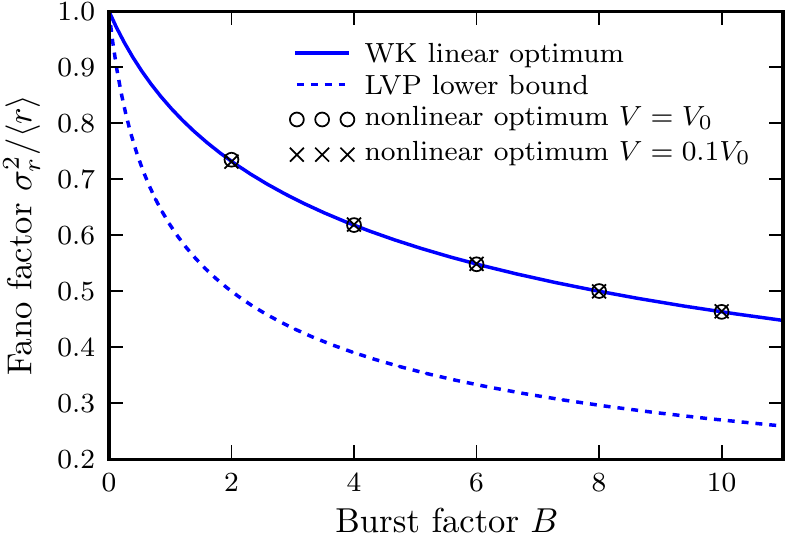}
\caption{The Fano factor $\sigma_r^2/\langle r \rangle$ as a function
  of burst ratio $B$.  The solid curve is the optimal result predicted
  by the WK linear theory, and the dashed curve is the rigorous lower
  bound derived by LVP~\cite{Lestas10}.  Symbols show numerical
  optimization for the generalized nonlinear TetR feedback system
  (Eq.~\eqref{nl3}) at two volumes, $V=V_0$ and
  $V=0.1V_0$.}\label{nonl}
\end{figure}

The semi-quantitative agreement between the linearized theory and the simulation
results displayed in Fig.~\ref{sim} still leaves open the possibility that some type of nonlinear,
discrete filter, not described by the experimentally fitted parameters
of the TetR gene network, could perform better than the WK optimum at
sufficiently small volumes.  Fig.~\ref{nonl} plots both the WK value
for the Fano factor (solid curve) and the rigorous lower bound of LVP
(dashed curve) as a function of $B$ (Eq.~\eqref{eq11}).  The above
question can be posed as follows: is it possible to achieve a Fano
factor that falls between the two curves by taking advantage of
nonlinearity and discreteness?  Ideally, one should do an optimization
over all possible nonlinear regulatory functions that could describe
feedback between the TetR protein and mRNA.  In full generality, such
an optimization appears intractable, but one can tackle a limited version
of the nonlinear optimization.  We will confine ourselves to Hill-like
regulatory functions, which describe the experimental behavior of many
cellular systems~\cite{Alon}, and explore whether it is possible to
find any scenario where this type of nonlinear feedback outperforms
the linear WK optimum.  We consider the following generalized TetR
feedback loop:
\begin{equation}\label{nl3}
\begin{split}
\dot{r}(t) &= -\gamma_r  r(t) + K_r(p(t)),\\
\dot{p}(t) &= -\gamma_p  p(t) - \Gamma_p(p(t)) + \kappa_p r(t),
\end{split}
\end{equation}
with two Hill-like regulatory functions,
\begin{equation}\label{nl4}
K_r(p) = \frac{A_1 \theta_1^{n_1}}{\theta_1^{n_1}+p^{n_1}}, \qquad 
\Gamma_p(p) = \frac{A_2 p^{n_2}}{\theta_2^{n_2}+p^{n_2}},
\end{equation}
involving arbitrary non-negative parameters $A_i$, $n_i$, $\theta_i$,
$i=1,2$.  The original TetR system (Eq.~\eqref{eq12}) is a special
case of the equations above with:
\begin{equation}\label{nl4b}
\begin{split}
A_1 &= \kappa_0,\quad n_1 = n,\quad \theta_1 = \theta,\quad  A_2 = \beta \Phi A,\\
n_2 &= 1,\quad \theta_2 =\gamma_a.
\end{split}
\end{equation}
The production function $K_r(p)$ is a monotonically decreasing
function of $p$, as is expected for negative feedback, while
$\Gamma_p(p)$ is monotonically increasing, a generalization of some
regulatory network which effectively removes the TetR protein from the
feedback loop (the role played by ATc binding in the experimental
system).  With these monotonicity constraints, there is always only
one steady-state solution $\bar{r}$ and $\bar{p}$ to Eq.~\eqref{nl3}.

The optimization consists of searching for $K_r(p)$ and $\Gamma_p(p)$
that minimize the Fano factor $\sigma_r^2/\langle r\rangle$.  The
following quantities are fixed during the search: the degradation
rates $\gamma_r$, $\gamma_p$, the $P$ production rate $\kappa_p$ (or
equivalently the burst ratio $B = \kappa_p /\gamma_r$), and the steady
state values $\bar{r}$, $\bar{p}$.  Note that in the general nonlinear
case, the steady state values do not necessarily coincide with the
mean values $\langle r \rangle$, $\langle p \rangle$, since the
equilibrium distributions are generally asymmetric with respect to the
steady state.  Fixing $\bar{r}$ and $\bar{p}$ during the optimization
is one way to set an overall copy number scale, to investigate the
role of discreteness.  It turns out that the optimization results
described below end up being independent of $\bar{r}$ and $\bar{p}$.
In terms of the Hill function parameters, fixing $\bar{r}$ and
$\bar{p}$ means setting $A_1$ and $A_2$ to the following values,
\begin{equation}\label{nl5}
\begin{split}
A_1 &= \theta_1^{-n_1}\gamma_r \bar{r} (\theta_1^{n_1}+\bar{p}^{n_1}),\\
A_2 &= \bar{p}^{-n_2} (\gamma_p \bar{p}-\kappa_p \bar{r})(\theta_2^{n_2}+\bar{p}^{n_2}).
\end{split}
\end{equation}
Thus the goal of optimization is to minimize $\sigma_r^2/\langle r
\rangle$ over the four remaining free parameters: $n_1$, $\theta_1$,
$n_2$, $\theta_2$.

In order to carry out this minimization, one needs an efficient
procedure to calculate $\sigma_r^2/\langle r \rangle$ from
Eq.~\eqref{nl3}, keeping both the full nonlinearity of the dynamical
system, and the discreteness of the $r(t)$ and $p(t)$ populations.
The system can always be simulated through the Gillespie
algorithm~\cite{Gillespie77}, and accurate estimates of $\langle r
\rangle$ and $\sigma_r^2$ determined from sufficiently long
trajectories.  However this approach is too slow for searching over
the four-dimensional parameter space, since each distinct set of
parameters would require a separate long simulation run.  An
equivalent, faster alternative is to directly solve the system's
master equation for the steady state probability distribution, which
then yields $\langle r \rangle$ and $\sigma_r^2$.  The joint
probability distribution $P_{r,p}(t)$ of finding $r$ mRNAs and $p$
proteins at time $t$ is governed by the master equation,
\begin{equation}\label{nl6}
\begin{split}
&\frac{\partial}{\partial t} P_{r,p}\\
&\qquad = \gamma_r \left[(r+1)P_{r+1,p} - r P_{r,p}\right]
+ K_r(p) \left[P_{r-1,p} - P_{r,p}\right]\\
&\qquad + \gamma_p\left[(p+1) P_{r,p+1} -  p P_{r,p}\right]+ \left[\Gamma_p(p+1) P_{r,p+1} \right.\\
&\qquad \left.-  \Gamma_p(p) P_{r,p}\right]+ \kappa_p r \left[P_{r,p-1}-P_{r,p}\right].  
\end{split} 
\end{equation}
The steady state distribution $P^s_{r,p}$ is the solution obtained by
setting to zero the right-hand side of the above equation, which we
denote ${\cal R}_{r,p}$:
\begin{equation}\label{nl7}
\begin{split}
0 &= {\cal R}_{r,p}\\
& \equiv  \gamma_r \left[(r+1)P^s_{r+1,p} -
  r P^s_{r,p}\right] + K_r(p) \left[P^s_{r-1,p} - P^s_{r,p}\right]\\
& +
\gamma_p\left[(p+1) P^s_{r,p+1} - p P^s_{r,p}\right]+
\left[\Gamma_p(p+1) P^s_{r,p+1}\right.\\
&\left. - \Gamma_p(p) P^s_{r,p}\right]+ \kappa_p
r \left[P^s_{r,p-1}-P^s_{r,p}\right].
\end{split} 
\end{equation}
The result is linear in the components $P^s_{r,p}$ for various $r$ and
$p$, and thus the set $ \{{\cal R}_{rp} =0\}$ for $r=0,1,\ldots$ and
$p=0,1,\ldots$, constitutes a linear system of equations for
$P^s_{r,p}$.  The master equation can be solved by spectral methods,
which are generally more efficient than brute force Gillespie
simulations~\cite{Mugler09PRE}.  However we use a different approach,
described below, to solve Eq.~\eqref{nl7}, which is sufficiently fast
for our numerical optimization purposes.  Since $r$ and $p$ can take
on any integer values between $0$ and $\infty$, we truncate the system
to focus only on the non-negligible $P^s_{r,p}$, in other words
$(r,p)$ within several standard deviations of the mean
$(\langle r \rangle, \langle p \rangle)$.  Specifically, we keep only
those equations ${\cal R}_{rp} =0$ which involve
$r_\text{min} \le r \le r_\text{max}$ and
$p_\text{min} \le p \le p_\text{max}$.  The largest truncation range
required for accurate results was $r_\text{max} - r_\text{min} = 100$
and $p_\text{max} - p_\text{min}=50$.  All $P^s_{r,p}$ outside the
range which appear in the truncated system of equations are set to a
positive constant $\epsilon >0$.  (The precise value of $\epsilon$ is
unimportant since the distribution is subsequently normalized, and the
truncation range is chosen large enough so that the boundary condition
does not significantly affect the outcome.)  The resulting finite
linear system, which is sparse, can be efficiently solved using an
unsymmetric-pattern multifrontal algorithm~\cite{Davis04}.  Knowing
$P^s_{r,p}$, we then directly calculate the moments of the
distribution to find $\langle r \rangle$ and $\sigma_r^2$.  The
numerical accuracy of the procedure is verified by comparison to
Gillespie simulation results.

In order to set a starting point for each round of nonlinear
optimization, we use the following initialization procedure: we take
the original TetR system at a given volume $V$ and burst ratio $B$
(fixing the Hill function parameters according to Eq.~\eqref{nl4b})
and find the ATc concentration $A_\text{min}$ where
$\sigma_r^2/\langle r \rangle$ is smallest, evaluating the Fano factor
using the linear solver described above.  The $\bar{r}$ and $\bar{p}$
at this concentration are then chosen to be fixed constants for
the nonlinear optimization, where we vary the parameters $n_1$,
$\theta_1$, $n_2$, $\theta_2$ from the initial values given by
Eq.~\eqref{nl4b} to minimize $\sigma_r^2/\langle r \rangle$.  The
minimization is carried out using Brent's principal axis
method~\cite{Brent02}, which is feasible due to the fast evaluation of
$\langle r \rangle$ and $\sigma_r^2$ at each different parameter set
through the linear solver.

\begin{figure*}
\centering\includegraphics[width=\textwidth]{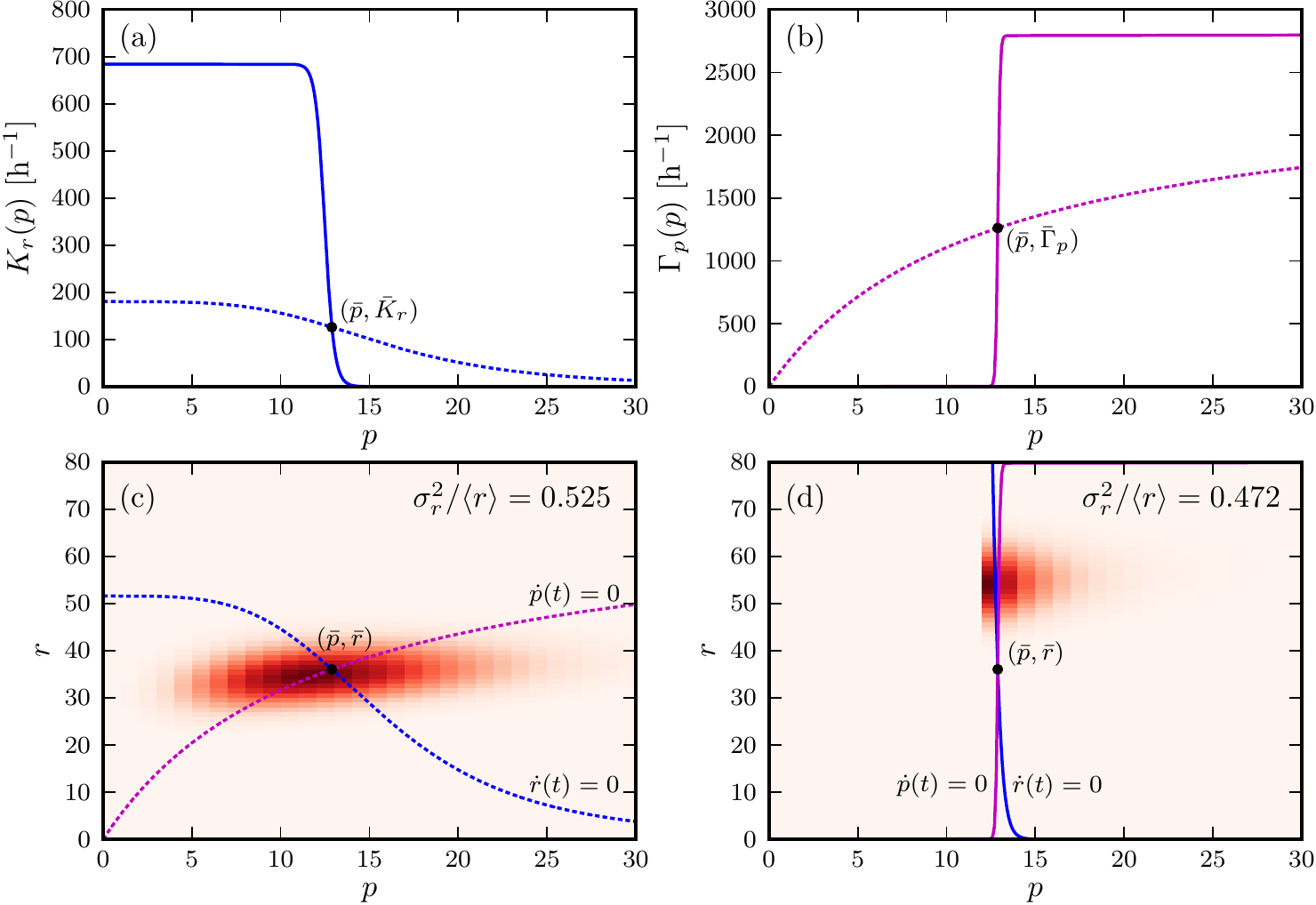}
\caption{Results for numerical optimization of the generalized nonlinear TetR
  feedback system of Eq.~\eqref{nl3}, with starting parameters $B=10$
  and $V=V_0$. (a) The mRNA production regulation function $K_r(p)$ in
  its initial form before optimization (dashed curve), and after
  several steps of the minimization algorithm (solid curve).  (b)
  Similar to (a), but showing the protein degradation function
  $\Gamma_p(p)$.  (c) Heat map of the steady-state probability
  distribution $P^s_{r,p}$ before optimization, corresponding to
  regulation governed by the dashed curves in the top panels.  The
  nullclines $\dot{r}(t) =0$ and $\dot{p}(t)=0$ are superimposed.  (d)
  Similar to (c), but after several steps of the minimization
  algorithm, corresponding to regulation governed by the solid curves
  in the top panels.}\label{nlfig1}
\end{figure*}

Fig.~\ref{nlfig1} shows results of a typical minimization run, where
the initial system is at volume $V=V_0$ with $B=10$, with a
corresponding $A_\text{min} = 50$ ng/mL.  The dashed lines in
Fig.~\ref{nlfig1}(a) and (b) show the Hill functions $K_r(p)$ and
$\Gamma_p(p)$ of the original TetR system at these parameter values,
and the heat map in Fig.~\ref{nlfig1}(c) represents the associated
steady-state probability distribution $P^s_{r,p}$.  The dashed lines
superimposed on the heat map are the loci of solutions to
$\dot{r}(t)=0$ and $\dot{p}(t)=0$ (the right-hand sides of
Eq.~\eqref{nl3} set to zero), which intersect at the steady state
$(\bar{r},\bar{p})$.  The Fano factor for this distribution, which
represents the best the TetR system can perform given the
experimentally fitted parameters, is
$\sigma^2_r/\langle r \rangle = 0.525$.  This is  above the
linear WK optimum for $B=10$, $2/(1+\sqrt{1+B}) = 0.463$, and
significantly larger than the rigorous LVP lower bound of
$2/(1+\sqrt{1+4B}) = 0.270$.  Once we relax the experimental
constraints, and carry out the numerical minimization, the Fano factor
decreases.  The solid lines in Fig.~\ref{nlfig1}(a) and (b) show
$K_r(p)$ and $\Gamma_p(p)$ after several steps of the minimization
algorithm, and Fig.~\ref{nlfig1}(d) shows the corresponding
$P^s_{r,p}$.  The Hill functions have become very steep steps around
$\bar{p}$, while the average of the distribution $\langle r \rangle$
has been pushed above $\bar{r}$.  The probabilities $P^s_{r,p}$ for
$p < p_0$ become negligible, where $p_0 \equiv \lfloor \bar{p}\rfloor$
is the largest integer value below $\bar{p}$.  For $p>p_0$,
$P^s_{r,p}$ rapidly decay to zero. The Fano factor,
$\sigma^2_{r}/\langle r \rangle = 0.472$, approaches closer to the
linear WK optimum, but is still above it.  If we allow the
minimization to proceed, these trends continue: at each iteration the
Hill functions get steeper, $\langle r \rangle$ increases, $P^s_{r,p}$
for $p < p_0$ tends to zero, and $\sigma^2_{r}/\langle r \rangle$
approaches arbitrarily close to the linear WK optimum from above.

In fact, the same behavior is seen irrespective of the volume $V$ and
burst ratio $B$ used to define the initial point of the optimization.
Fig.~\ref{nonl} shows the results of nonlinear optimization for
$B=2-10$ at two volumes, $V=V_0$ and $V=0.1V_0$.  Even for the
smallest volume, the nonlinear optimization results can get
arbitrarily close to the WK optimum, but never do better.  No
generalized nonlinear system based on Hill function regulation brings
us close to the theoretically possible LVP lower bound.  This overall
conclusion holds even when we change the functional form for the
generalized feedback.  We tried two alternatives: (i) using sigmoidal
(logistic) functions instead of Hill functions; (ii) expanding
$K_r(p)$ and $\Gamma_p(p)$ in a Taylor series around $\bar{p}$,
truncating after the third order term, and minimizing with respect to
the Taylor coefficients.  In both cases numerical minimization of the
Fano factor led to similar step-like behavior for $K_r(p)$ and
$\Gamma_p(p)$, and the Fano factor tended to WK optimum from above.

From the $P^s_{r,p}$ distribution in Fig.~\ref{nlfig1}(d) we see that
the step-function limit leads to a system which is highly nonlinear
along the $p$ axis: in fact the gene network spends most of its time
at $p=p_0$, just below the sudden change in regulation due to the
steep Hill functions, and $p > p_0$ just above the sudden regulatory
change.  The feedback on the TetR mRNA population is mediated by $p$
fluctuations between the two regimes, resulting in threshold-like
regulatory behavior.  Remarkably, despite this discrete, nonlinear
character, the network can still approach the efficiency of an optimal
WK linear filter.  To gain a deeper understanding of how the step-like
regulation can match WK optimality, we used the numerical optimization
results described above to posit a limiting form of the nonlinear gene
network that can be solved analytically (details in Appendix C).  The
analytic results explicitly show that we can asymptotically approach
the WK optimum behavior from above, even in systems where the protein
copy numbers are very small.  Thus at least for a two-component
TetR-like system regulated by biologically-realistic Hill functions,
the constraint derived from the WK theory has a broader validity than
one would guess from the underlying continuum, linear assumptions.  It
thus becomes an interesting and a non-trivial problem, left for future
studies, to find an example of a gene network where the rigorous lower
bound of LVP could be directly achieved.

\subsection*{Realizing optimality under the influence of extrinsic noise}

Extrinsic noise is ubiquitous and hence must also be considered in any
effective description of the control network.  Inevitably, certain
cellular components are not explicitly included in such a description,
which in our case study could include RNA polymerase, ribosomes, and
transcription factors that bind to the same promoter.  Each of these
components have their own stochastic characteristics and may
contribute noise to a smaller or greater extent.  Particularly for
eukaryotes like yeast, the extrinsic noise contribution may be
significantly larger than the intrinsic
component~\cite{Raser04,Volfson06}.  We adopt a simple model for the
extrinsic noise based on earlier approaches~\cite{Austin06,Cox06},
which assume that it is band-limited at a low frequency $\tau_e^{-1}$,
where $\tau_e$ is on the order of the cell growth time scale.  The
justification is that higher frequency contributions to the extrinsic
noise are filtered out by the gene circuits associated with its
sources.  This idea is consistent with the experimental observation of
extrinsic noise in protein production in {\it E. coli}, which found
long autocorrelation times for the extrinsic noise on the order of the
cell cycle period~\cite{Rosenfeld05}.

\begin{figure}[t]
\centering\includegraphics[width=\columnwidth]{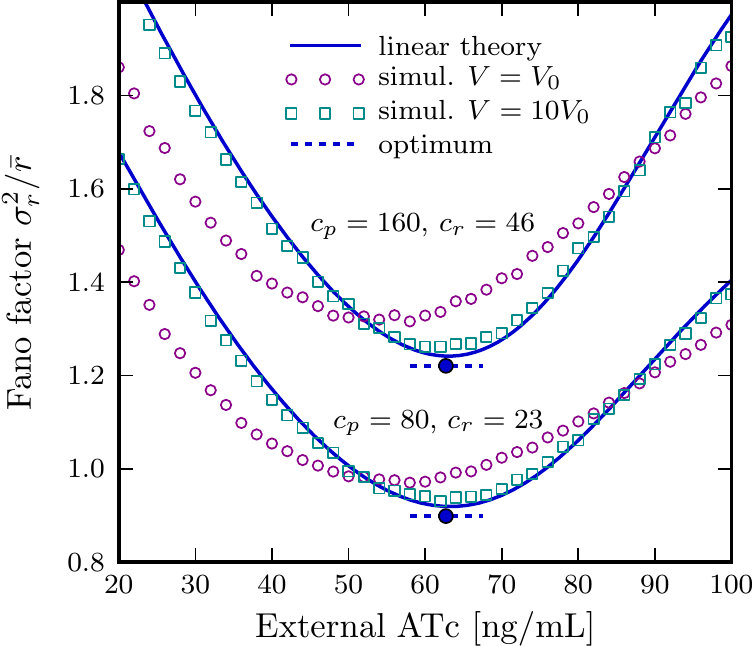}
\caption{Comparison of simulation and theory results based on the
  dynamical model (Eq.~\eqref{eq12}) of the yeast synthetic gene
  circuit~\cite{Nevozhay09}, in the presence of extrinsic noise given
  by Eq.~\eqref{eq14}.  All quantities are plotted as a function of
  extracellular ATc concentration $A$ for the burst ratio $B=5$.  Each
  set of curves shows the Fano factor $\sigma^2_r/\bar{r}$, as
  predicted by the linear filter theory (solid lines), versus
  stochastic numerical simulations at two different volumes,
  $V = V_0 = 60$ fL (circles) and $V = 10V_0$ (squares).  The two sets
  correspond to noise magnitudes $c_p = 80$, $c_r = 23$ and
  $c_p = 160$, $c_r = 46$.  In both cases $c_r$ and $c_p$ are related
  through the condition in Eq.~\eqref{si17}, and the minimal Fano
  factor predicted by WK filter theory (horizontal dashed lines) is
  modified as shown in Eq.~\eqref{eq17}.  The system can be tuned to
  approach optimality near a particular $A_{\rm opt}$ obtained by the
  condition $G_{rp} = {\cal G}^{\rm opt}_{rp}$ (filled
  circles).}\label{extr}
\end{figure}

For the TetR system, our theory is extended to the extrinsic noise
case in Appendix D, with the results illustrated in Fig.~\ref{extr}.
The outcome is that a given TetR gene circuit, tuned appropriately
such that $A = A_\text{opt}$, can act as a WK filter for an entire
family of extrinsic noise scenarios.  A single set of parameters can
approximately represent the optimal solution for a variety of
extrinsic inputs.  This makes the WK concept a versatile design tool
for noise suppression in biological systems: the same control network
can act with maximum efficiency in a variety of different contexts.
It is possible that the requirement of adaptability to a wide range of
conditions has resulted in the evolution of control networks acting as
WK filters.  It remains to be seen whether nature has exploited this
feature {\it in vivo}.

\section*{Conclusion}

The TetR feedback loop is a concrete example of how a WK filter can be
implemented in a gene network driven by a complex set of biochemical
reaction rates, but the overall approach outlined here has far
reaching implications, thus highlighting the appeal of engineering
paradigms in biology~\cite{Csete02Science}.  With the entire network
complexity encoded in a handful of response functions, we can derive
fundamental limits and design principles governing biological
regulation.  The key step is to map the linear response picture onto a
signal estimation problem, whose solution is given by WK theory.  This
idea allows us to predict the dynamic properties of the feedback
pathway required to optimally filter noise in a broad class of
negative feedback circuits.  As already demonstrated in earlier
works~\cite{Hinczewski2014,Becker2015}, the mapping, and the potential
utility of the WK approach, is not unique to the negative feedback
loop.  Another important byproduct of the theory is that the behavior
of gene circuits away from optimality can also be predicted. In this
sense, our practical approach goes beyond just obtaining rigorous
bounds, and allows us to characterize how close or far gene networks
are from optimality for biologically relevant parameters.

We have derived response functions by linearizing a minimal model
extracted from experimental observations, but it is also possible to
directly apply small perturbations to a system, and measure the
resulting time-dependent changes in populations of species.  Recently,
the yeast hyperosmolar signaling pathway has been probed by
perturbations in the form of salt
shocks~\cite{Mettetal08,Hersen08,Muzzey09}.  Despite the underlying
complex nonlinear network, the details of which are not completely
characterized, a linear response description quantitatively captures
the frequency-dependent behavior of the pathway over a wide range of
inputs.  {\it E. Coli} chemotaxis signaling also exhibits a linear
regime~\cite{Park10}, where the fluctuation-dissipation relationship
between the system's unperturbed behavior and its reaction to external
stimuli has been explicitly verified.

Linear response functions can thus become a fundamental tool in
analyzing biochemical circuits, analogous to their established role in
control engineering and signal processing.  More extensive
experimental measurements will be critical in this effort, in order to
ascertain how varied  the response relationships between regulatory
components are in nature.  Once we understand the essential dynamical
building blocks out of which complex biological function is realized,
we can map out the hidden constraints that control the behavior of
living systems.

\begin{acknowledgments}
  This work was done while the authors were in the Institute for
  Physical Sciences and Technology in the University of Maryland,
  College Park. We are grateful to C. G\"uven, G. Reddy, Z. Zhang, and
  P. Zhuravlev for useful discussions. This work was supported by a
  grant from the National Science Foundation (CHE 13-61946).
\end{acknowledgments}

\bibliography{signal_refs}

\appendix

\section{Derivation of the optimal WK filter}

In this section we derive Eqs.~\eqref{eq6} and \eqref{eq7} in the main
text.  They describe the output variance $\sigma_r^2$ =
$[\overline{(\delta r)^2}]$ and the linear filter
$H_{\rm opt}(\omega)$ that minimizes $\sigma_r^2$, which are the main
quantities in the Wiener-Kolmogorov theory.

\subsection*{Output variance $\sigma_r^2$ in terms of signal and noise power spectra $P_s(\omega)$ and $P_n(\omega)$}

From Eq.~\eqref{eq4}, which defines the signal $s(\omega)$
and estimate $\tilde{s}(\omega)$ in the frequency domain, the Fourier
transformed output $\delta r(\omega)$ for any $H(\omega)$ can be
rewritten as,
\begin{equation}\label{wk1}
\delta r(\omega) = s(\omega) - \tilde{s}(\omega) = (1-H(\omega))s(\omega) - H(\omega) n(\omega). 
\end{equation}
In the time domain, $s(\omega) = -n_r(\omega)/(G_{rr}(\omega) +
i\omega)$, is a convolution of the noise function $n_r(t)$, and
$n(\omega) = n_p(\omega)/G_{pr}(\omega)$ is a convolution of $n_p(t)$.
So long as the noise functions $n_r(t)$ and $n_p(t)$ are uncorrelated,
$s(t)$ and $n(t)$ are also uncorrelated, so the frequency domain
average $\overline{s(\omega)n(\omega^\prime)} =0$.  (The theory can
also be generalized to correlated noise sources, but for
simplicity we consider only the uncorrelated case.)  As a result, the
correlation $\overline{\delta r(\omega) \delta r(\omega^\prime)}$,
related to the output power spectrum $P_{\delta r}(\omega)$, can be
written in terms of $P_s(\omega)$ and $P_n(\omega)$, the individual
power spectra of the signal and noise:
\begin{equation}\label{wk2}
\begin{split}
&\overline{\delta r(\omega) \delta r(\omega^\prime)}\\
&\quad= (1-H(\omega))(1-H(\omega^\prime)) \overline{s(\omega)s(\omega^\prime)}\\
&\quad \qquad  +H(\omega)H(\omega^\prime) \overline{n(\omega)n(\omega^\prime)}\\
&\quad = 2\pi \left[ |H(\omega)-1|^2 P_s(\omega) + |H(\omega)|^2 P_n(\omega)\right] \delta(\omega + \omega^\prime)\\
&\quad \equiv 2\pi P_{\delta r}(\omega) \delta(\omega+\omega^\prime),
\end{split}
\end{equation}
In the above equation we have used the definition of the power spectrum,
i.e. $\overline{s(\omega)s(\omega^\prime)} \equiv 2\pi P_s(\omega)
\delta(\omega+\omega^\prime)$, and the relation $H(-\omega) =
H^\ast(\omega)$ since $H(\omega)$ is the Fourier transform of a real
function $H(t)$.  The power spectrum $P_{\delta r}(\omega)$ is the
Fourier transform of the time autocorrelation function
$\overline{\delta r(t)\delta r(0)}$:
\begin{equation}\label{wk3}
\overline{\delta r(t)\delta r(0)} = \int_{-\infty}^{\infty} \frac{d\omega}{2\pi} P_{\delta r}(\omega) e^{-i \omega t}.
\end{equation}
At $t=0$, the autocorrelation function gives us the variance
$\sigma_r^2$:
\begin{equation}\label{wk4}
\begin{split}
\sigma_r^2 &= \overline{(\delta r(0))^2}\\
 &= \int_{-\infty}^{\infty} \frac{d\omega}{2\pi} P_{\delta r}(\omega)\\
&= \int_{-\infty}^{\infty} \frac{d\omega}{2\pi}\, \left[|H(\omega)|^2 P_n(\omega) + |H(\omega)-1|^2 P_s(\omega) \right],
\end{split}
\end{equation}
which is Eq.~\eqref{eq6} in the main text.

\subsection*{Minimizing $\sigma_r^2$ over all causal $H(\omega)$ yields the optimal WK filter $H_{\rm opt}(\omega)$}

The convolution of the filter function $H(t)$ on the corrupted signal
$s(t) + n(t)$ must satisfy causality.  The filter can only operate on
the past history of $s(t) + n(t)$, so $H(t) = 0$ for $t<0$.  In the
frequency domain, enforcing causality restricts $H(\omega)$ to have
certain general properties as a function of complex
$\omega$~\cite{ChaikinLubensky}: it can have no poles or zeros in the
upper half-plane ${\rm Im}\,\omega >0$.  Equivalently, the real and
imaginary parts of $H(\omega)$ evaluated at real $\omega$ must satisfy
the well-known Kramers-Kronig relation:
\begin{equation}\label{wk5}
{\rm Re}\,H(\omega) = \frac{1}{\pi} {\cal P} \int_{-\infty}^\infty d\omega^\prime \frac{{\rm Im}\,H(\omega^\prime)}{\omega^\prime - \omega},
\end{equation}
where ${\cal P}$ is the Cauchy principal value of the integral.  The
goal of WK optimization is to minimize $\sigma_r^2$ in
Eq.~\eqref{wk4} over all possible causal functions $H(\omega)$,
given the power spectra $P_s(\omega)$ and $P_n(\omega)$.

Assume such an optimum $H_{\rm opt}(\omega)$ exists, with
the corresponding minimal variance $\sigma^2_{r,{\rm opt}}$.  Let us add a small
perturbation, $H(\omega) = H_{\rm opt}(\omega) + \delta H(\omega)$, where
$\delta H(\omega)$ is also a causal function of complex $\omega$.
From Eq.~\eqref{wk4}, the resulting variance change $\delta \sigma_r^2
= \sigma^2_r - \sigma^2_{r,{\rm opt}}$, to lowest order in $\delta
H(\omega)$, is:
\begin{equation}\label{wk6}
\begin{split}
\delta \sigma^2_r &= \int_{-\infty}^\infty d\omega\, 2\, {\rm Re}\bigl[ \left\{(H_{\rm opt}(\omega) -1)P_s(\omega)\right.\\
&\qquad  + \left. H_{\rm opt}(\omega)P_n(\omega)\right\}\delta H^\ast(\omega)\bigr]\\
&= \int_{-\infty}^\infty d\omega\, 2\, {\rm Re}\bigl[{\cal F}_{\rm opt}(\omega) \delta H^\ast(\omega) \bigr],
\end{split}
\end{equation}
where
\begin{equation}\label{wk6b}
{\cal F}_{\rm opt}(\omega) \equiv (H_{\rm opt}(\omega)
-1)P_s(\omega) + H_{\rm opt}(\omega)P_n(\omega).
\end{equation}
  For $H_{\rm opt}(\omega)$ to be the WK optimum, $\delta \sigma_r^2$
  in Eq.~\eqref{wk6} must be zero for any causal perturbation $\delta
  H(\omega)$.

  Out of all possible causal perturbations, we will focus on one with
  the specific form:
\begin{equation}\label{wk7}
\delta H(\omega) = \frac{A}{\epsilon - i (\omega - \omega_0)},
\end{equation}
where ${\rm Im}\,\omega_0 =0$ and $A, \epsilon >0$.  It has no zeros,
and the only pole, $\omega = \omega_0 - i\epsilon$, is in the 
lower half-plane, so $\delta H(\omega)$ is causal.  We will be
interested in the limit as this pole approaches the real axis,
$\epsilon \to 0^+$, where the real and imaginary parts of $\delta
H(\omega)$ are,
\begin{equation}\label{wk8}
\begin{split}
{\rm Re}\,\delta H(\omega) &= \frac{A \epsilon}{\epsilon^2 + (\omega-\omega_0)^2} \to A \pi \delta (\omega - \omega_0),\\
{\rm Im}\,\delta H(\omega) &= \frac{A (\omega -\omega_0)}{\epsilon^2 + (\omega-\omega_0)^2} \to \frac{A}{\omega - \omega_0}.
\end{split}
\end{equation}
Substituting these into Eq.~\eqref{wk6} for $\delta \sigma_r^2$, we
find that the optimality condition $\delta \sigma_r^2 = 0$ implies the
following relation between the real and imaginary parts of ${\cal
  F}_{\rm opt}(\omega)$:
\begin{equation}\label{wk9}
{\rm Re}\,{\cal F}_{\rm opt}(\omega_0) = -\frac{1}{\pi} {\cal P} \int_{-\infty}^{\infty} d\omega \frac{{\rm Im}\,{\cal F}_{\rm opt}(\omega)}{\omega - \omega_0}.
\end{equation}
This has the same form as the Kramers-Kronig relation in
Eq.~\eqref{wk5}, with the important difference of a minus sign in
front.  Consequently, ${\cal F}_{\rm opt}(\omega)$ must be {\it
  anticausal}, which we define as a function with no poles or zeros in
the lower complex $\omega$ half-plane.

In order to use this result to derive a solution for $H_{\rm
  opt}(\omega)$, we define two types of decompositions, described
briefly in the main text.  In practice, all the frequency domain power
spectral density and filter functions we work with in the linear
response formalism are meromorphic functions over the complex $\omega$
plane.  Any meromorphic function $F(\omega)$ can be written as a
partial fraction expansion of the form $F(\omega) = \sum_{n,k}
c_{ik}/(\omega - \omega_n)^k$, where $\{ \omega_n \}$ is the set of
poles of $F(\omega)$, and $c_{ik}$ are constants.  Most generally, the
expansion could include a polynomial term, but the functions
$F(\omega)$ we encounter have well-defined inverse Fourier transforms,
which require $|F(\omega)| \to 0$ as $|\omega| \to \infty$ (decay at
least as fast as $1/|\omega|$).  Thus, all the terms in the expansion
are of the form $c_{ik}/(\omega - \omega_n)^k$, and we can segregate
them according to whether the pole $\omega_n$ is in the upper half
plane.  The causal part $\{F(\omega)\}_{c}$ is defined as all those
terms where $\omega_n$ is not in the upper half plane, and the
anticausal part $\{F(\omega)\}_{ac}$ contains the remaining terms in
the expansion.  The overall function $F(\omega) = \{F(\omega)\}_{c} +
\{F(\omega)\}_{ac}$.

The second type of decomposition, an example of Wiener-Hopf
factorization~\cite{Wiener49}, concerns power spectral density
functions like $P_y(\omega)$, which are meromorphic and also
real-valued on the real $\omega$ axis.  Let us factor $P_y(\omega)$ as
the product of two meromorphic functions, $P_y(\omega) = P_y^c(\omega)
R_y^{ac}(\omega)$.  The function $P_y^c(\omega)$ contains all the
zeros and poles in $P_y(\omega)$ which are not in the upper half
plane.  Such a decomposition is always possible, since a meromorphic
function can always be written as a ratio of two holomorphic
functions. Hence, the numerator and denominator of $P_y(\omega)$ can
be decomposed individually into a product of elementary factors by the
Weierstrass factorization theorem, with each factor containing a
single zero.  Because $P_y(\omega)$ is real for real $\omega$, so
$P_y(\omega)^\ast = P_y(\omega)$ when ${\rm Im}\,\omega =0$.  Thus,
$P_y^c(\omega)^\ast R_y^{ac}(\omega)^\ast = P_y^c(\omega)
R_y^{ac}(\omega)$. Since $P_y^c(\omega)^\ast$ for real $\omega$ has
all its zeros and poles in the upper half plane, we must have
$P_y^c(\omega)^\ast \propto R_y^{ac}(\omega)$, and similarly
$R_y^{ac}(\omega)^\ast \propto P_y^c(\omega)$.  By appropriately
absorbing an overall constant into $P_y^c(\omega)$, we can factor
$P_y(\omega)$ as $P_y(\omega) = P_y^c(\omega) P_y^c(\omega)^\ast =
|P_y^c(\omega)|^2$.

With these decompositions defined, we return now to the condition in
Eq.~\eqref{wk9}, which shows that ${\cal F}_{\rm opt}(\omega)$ is
anticausal.  Thus, its causal part in the additive decomposition must
be zero, $\{ {\cal F}_{\rm opt}(\omega) \}_c = 0$.  From the
definition of ${\cal F}_{\rm opt}(\omega)$, Eq.~\eqref{wk6b}, it
follows that
\begin{equation}\label{wk10}
\{H_{\rm opt}(\omega)P_y(\omega)\}_c  =  \{P_s(\omega)\}_c,
\end{equation}
where $P_y(\omega) = P_s(\omega) + P_n(\omega)$ is the power spectrum
of the noise-corrupted signal $y(t) = s(t) + n(t)$.  Equivalently,
since we can substitute $\{F(\omega)\}_c = F(\omega) -
\{F(\omega)\}_{ac}$ for any $F(\omega)$, the optimality condition can
be rewritten as:
\begin{equation}\label{wk11}
\begin{split}
&H_{\rm opt}(\omega)P_y(\omega) - \{H_{\rm opt}(\omega)P_y(\omega)\}_{ac}\\
&\qquad  = P_s(\omega) - \{P_s(\omega)\}_{ac}.
\end{split}
\end{equation}
Divide both sides of Eq.~\eqref{wk11} by $P_y^c(\omega)^\ast$, and
then take the causal additive part $\{ \cdot \}_c$ of both sides.  The
result is:
\begin{equation}\label{wk12}
\begin{split}
&\{ H_{\rm opt}(\omega)P_y^c(\omega)\}_c - \left\{\frac{\{H_{\rm opt}(\omega)P_y(\omega)\}_{ac}}{P_y^c(\omega)^\ast}\right\}_c\\
&\qquad  = \left\{\frac{P_s(\omega)}{P_y^c(\omega)^\ast}\right\}_c - \left\{ \frac{ \{P_s(\omega)\}_{ac}}{P_y^c(\omega)^\ast}\right\}_{c}.
\end{split}
\end{equation}
The second terms on both the left and right hand sides are the causal
parts of a ratio between two anticausal functions.  Since a ratio of
anticausal functions is also anticausal, these terms are zero.  On the
left hand side the first term $\{
H_{\rm opt}(\omega)P_y^c(\omega)\}_c =
H_{\rm opt}(\omega)P_y^c(\omega)$, since $H_{\rm opt}(\omega)$ and
$P_y^c(\omega)$ are causal, and hence their product is also
causal. Making these simplifications, we can then solve for
$H_{\rm opt}(\omega)$ as:
\begin{equation}\label{wk13}
H_{\rm opt}(\omega) = \frac{1}{P_y^c(\omega)}\left\{ \frac{P_s(\omega)}{P_y^c(\omega)^\ast}\right\}_c,
\end{equation}
which is the optimal WK filter result shown as
Eq.~\eqref{eq7} in the main text.

\section{Linear response and noise filter analysis for a regulatory cascade}

As an example of how our theory generalizes to control networks with
multiple mediator species, we will consider the case where the
feedback loop consists of a regulatory cascade.  We will still
explicitly single out a target species $R$ and a mediator $P$, but now
the signaling pathway which communicates changes from $R$ to $P$ will
be more complicated, consisting of a cascade of $N$ species $U_j$,
$j=1,\ldots,N$, with populations $u_j$.  The production of the
$j^{th}$ species will depend on the population of the $(j-1)^{th}$
species (with $j=0$ corresponding to $R$), and $P$ will depend on the
last member of the cascade, $U_N$.  In terms of Fourier-transformed
fluctuations $\delta u_j$, the dynamical equations for the pathway
have the form:
\begin{equation}\label{c1}
\begin{split}
-i \omega \delta u_j(\omega) &= G_{u_j u_j}(\omega) \delta u_j(\omega) + G_{u_j u_{j-1}}(\omega) \delta u_{j-1}(\omega)\\
&\quad + n_{u_j}(\omega), \qquad j=1,\ldots,N.
\end{split}
\end{equation}
Thus the dynamics includes three parts: (i) the self-responses $G_{u_j
  u_j} $ which we can assume in the simplest case to be given by the
inverse decay lifetimes of the species, $G_{u_j u_j} =
-\tau_{u_j}^{-1}$; (ii) the cross-response terms $G_{u_j u_{j-1}}$
which describe how the $j$th member of the cascade is related to the
$(j-1)$th member; (iii) the stochastic noise terms $n_{u_j}$.  To
complete the description of the feedback loop, we specify the
equations for $R$ and $P$:
\begin{equation}\label{c2}
\begin{split}
-i \omega \delta r(\omega) &= G_{r r}(\omega) \delta r(\omega) + G_{r p}(\omega) \delta p(\omega) + n_{r}(\omega),\\
-i \omega \delta p(\omega) &= G_{p p}(\omega) \delta p(\omega) + G_{p u_{N}}(\omega) \delta u_{N}(\omega) + n_{p}(\omega).
\end{split}
\end{equation}
Instead of the simple cross-response $G_{pr}$ from $R$ to $P$, 
$P$ is influenced by the final species of the $U_j$ pathway through $G_{p
  u_{N}}$.

The regulatory cascade system described by Eqs.~\eqref{c1}-\eqref{c2}
can in fact be simplified extensively, by solving for the dynamics of
the mediator species $U_j$ and substituting the results into
Eq.~\eqref{c2}.  This yields equations for $R$ and $P$ which have the same form
as in the two-species case in the main text, but with an
effective cross-response function $G^{\rm eff}_{pr}(\omega)$ and
noise term $n^{\rm eff}_p(\omega)$,
\begin{equation}\label{c3}
\begin{split}
-i \omega \delta r(\omega) &= G_{r r}(\omega) \delta r(\omega) + G_{r p}(\omega) \delta p(\omega) + n_{r}(\omega)\\
-i \omega \delta p(\omega) &= G_{p p}(\omega) \delta p(\omega) + G_{p r}^{\rm eff}(\omega) \delta r(\omega) + n^{\rm eff}_{p}(\omega),
\end{split}
\end{equation}
where:
\begin{equation}\label{c4}
\begin{split}
G^{\rm eff}_{pr}(\omega) &= G_{p u_{N}}(\omega) \prod_{j=1}^N \frac{G_{u_j u_{j-1}}(\omega) \tau_{u_j}}{1-i \omega \tau_{u_j}},\\
n_p^{\rm eff}(\omega) &= n_p(\omega) \\
&+ G_{p u_{N}}(\omega) \sum_{k=1}^N \frac{n_{u_k}(\omega)}{G_{u_k u_{k-1}}(\omega)}\prod_{j=k}^N \frac{G_{u_j u_{j-1}}(\omega)\tau_{u_j}}{1-i\omega \tau_{u_j}}.
\end{split}
\end{equation}
In this effective two-species reduction of the full system, all the
stochastic effects of the mediators in the $U_i$ pathway enter in as
``extrinsic'' noise contributions to $n^{\rm eff}_p(\omega)$.  This
is a particular example that shows how extrinsic noise encapsulates the
stochastic influence of all the species that are not explicitly
specified in the dynamical equations.

The mapping of the two-species system onto the noise filter formalism,
and the calculation of the optimal filter, can be carried out by the
methods outlined in the main text.  While this in general results in a
more complicated problem than the simple example analyzed in the main
text, in one scenario the noise filter optimization problem for the
cascade is relatively straightforward: (i) we assume linear production
functions $k^+_{u_j}(t) = \kappa_{u_j} u_{j-1}(t)$ for all $U_j$, so
the cross-responses are constants in frequency space, $G_{u_j
  u_{j-1}}(\omega) \equiv \kappa_{u_j}$.  Similarly, the $P$
production function is $\kappa_p u_N(t)$, so $G_{p u_N}(\omega) \equiv
\kappa_p$.  (ii) We assume the decay timescales of all the cascade
species are negligible, $\tau_{u_j} \ll \tau_r$, so we can take the
limits $\tau_{u_j}\to 0$ in Eq.~\eqref{c4}.  However, the products
$\kappa_{u_j} \tau_{u_j}$ remain finite for all $j$, since from the
equilibrium conditions of the cascade (balance of production and
destruction), they are related to ratios of the steady-state populations
$\bar{u}_j$:
\begin{equation}\label{c5}
\kappa_{u_j} \tau_{u_j} = \frac{\bar{u}_j}{\bar{u}_{j-1}}.
\end{equation}
Hence rapid decay goes hand in hand with fast production.  This is the
same type of serial cascade analyzed in Ref.~\citenum{Lestas10}, where it
was shown to maximize information transfer along the pathway.  (iii)
Finally, we assume that each species in the original, full description
of the system is subject only to intrinsic noise, so the noise
functions are given by:
\begin{equation}\label{c6}
\begin{split}
n_r(\omega) &= \sqrt{2\bar{k}_r} \eta_r(\omega),\\
 n_p(\omega) &= \sqrt{2\kappa_p\bar{u}_N} \eta_p(\omega),\\
 n_{u_j}(\omega) &= \sqrt{2\kappa_{u_j}\bar{u}_{j-1}} \eta_{u_j}(\omega), 
\end{split}
\end{equation}
where the $\eta_\alpha(\omega)$ for different $\alpha$ are independent
Fourier-transformed Gaussian white noise functions.

With these assumptions the effective cross-response and noise
functions in Eq.~\eqref{c4} become:
\begin{equation}\label{c7}
\begin{split}
G^{\rm eff}_{pr}(\omega) &= \frac{B}{\tau_r},\\
n_p^{\rm eff}(\omega) &= n_p(\omega) + B \sum_{k=1}^N \frac{n_{u_k}(\omega)}{B_{u_k}},
\end{split}
\end{equation}
where the $P$ burst ratio $B \equiv \kappa_p \bar{u}_N \tau_r/\bar{r}$
is  analogous to $B$ in the main text, i.e. the average
number of $P$ molecules produced per $R$ during the time interval
$\tau_r$.  Similarly the burst ratio $B_{u_k} = \kappa_{u_k}
\bar{u}_{k-1} \tau_r/\bar{r}$ is the average number of $U_k$ molecules
produced per $R$ during $\tau_r$.

The resulting signal and noise power spectra within the filter formalism are:
\begin{equation}\label{c8}
P_s(\omega) = \frac{2 \bar{r} \tau_r}{1 + (\omega \tau_r)^2}, \quad P_n(\omega) = \frac{2\bar{r}\tau_r}{B_{\rm eff}},
\end{equation}
where:
\begin{equation}\label{c9}
B_{\rm eff} = \left[\frac{1}{B} + \sum_{k=1}^N \frac{1}{B_{u_k}} \right]^{-1}.
\end{equation}
Since the power spectra in Eq.~\eqref{c8} have the same form as
Eq.~\eqref{si1}, with $B$ replaced by $B_{\rm eff}$,
all the subsequent optimality results are identical, but expressed in
terms of the effective total burst ratio $B_{\rm eff}$ of the
signaling pathway.  This agrees with the effective burst ratio for the
cascade derived by the information theory approach in
Ref.~\citenum{Lestas10}, under the assumptions of rapid production/decay
outlined above.  Physically, this result implies that $B_{\rm eff}$
will be dominated by the smallest values among the $B$ and $B_{u_k}$.
Hence,  the efficiency of the noise filtration in the cascade is limited
by the weakest links.

\subsection*{Analytic limiting form of the generalized
  nonlinear feedback network }

We will use the numerical optimization results described in the main
text for the generalized nonlinear TetR feedback network
(Eq.~\eqref{nl3}) to derive a limiting form of the system that can be
solved analytically.  Since the optimization algorithm results in
steep step-like functions $K_r(p)$ and $\Gamma_p(p)$ with thresholds
at $\bar{p}$, let us assume that optimal limit for these Hill
functions looks like:
\begin{equation}\label{nl8}
K_r(p) = K_r^0 \Theta(\bar{p}-p), \quad \Gamma_p(p) = \Gamma_p^0 \Theta(p-\bar{p}),
\end{equation}
where the Heaviside step function $\Theta(x) = 0$ for $x<0$ and
$\Theta(x) = 1$ for $x>0$.  The plateau heights $K_r^0>0$ and
$\Gamma_p^0>0$ are assumed to be large, with a well defined ratio $\xi
\equiv K_r^0/\Gamma_p^0$ as $K_r^0, \Gamma_p^0 \to \infty$.  Since
$\Gamma_p^0 \gg \gamma_p$ and thus $\Gamma_p(p)$ acts as the dominant protein
degradation term, we will set $\gamma_p =0$ for simplicity.  (This has
negligible effect on the resulting $P^s_{r,p}$, particularly since
$\gamma_p^{-1} = 8.3$ h was already the longest
time scale in the system.)

Under these assumptions, we would like to find an analytical
steady-state probability distribution $P^s_{r,p}$ which satisfies
${\cal R}_{rp} = 0$ from Eq.~\eqref{nl7} for all $r,p\ge 0$.  We
cannot solve the system of equations directly, but we will introduce
an ansatz for $P^s_{r,p}$ and verify that it is a solution to
Eq.~\eqref{nl7}.  The first part of the ansatz is trivial: we assume
$P^s_{r,p} = 0$ for $p < p_0 = \lfloor \bar{p} \rfloor$.  This
satisfies ${\cal R}_{rp} = 0$ for $p<p_0$ exactly, regardless of the
values of $P^s_{r,p}$ at $p\ge p_0$.  To motivate the second part of
the ansatz, which covers the $p \ge p_0$ region, we need some more
information about the moments of the distribution.  This can be
gathered by defining the generating function,
\begin{equation}\label{nl9}
F(z_1,z_2) = \sum_{r=0}^\infty \sum_{p=p_0}^\infty  z_1^r z_2^{p-p_0} P^s_{r,p}.
\end{equation}
Summing the steady-state conditions ${\cal R}_{rp} = 0$ in
Eq.~\eqref{nl7} over all $r>0$, $p\ge p_0$, we obtain an equation that
can be expressed in terms of $F$:
\begin{equation}\label{nl10}
\begin{split}
&\gamma_r(1-z_1) F^{(1,0)}(z_1,z_2) + K_r^0 (z_1-1)F(z_1,0)\\
& + \Gamma_p^0(z_2^{-1}-1)\left[F(z_1,z_2)-F(z_1,0)\right]\\
&+ \kappa_p z_1 (z_2-1) F^{(1,0)}(z_1,z_2) =0,
\end{split}
\end{equation}
where $F^{(i,j)}(z_1,z_2) \equiv \partial_{z_1}^i \partial_{z_2}^j
F(z_1,z_2)$.  Taking the $z_1$ derivative of Eq.~\eqref{nl10}, and evaluating the result at $z_1=1$, $z_2=1$, gives:
\begin{equation}\label{nl11}
-\gamma_r F^{(1,0)}(1,1) + K_r^0 F(1,0) = 0.
\end{equation}
Similarly, differentiating Eq.~\eqref{nl10} with respect to $z_2$ yields:
\begin{equation}\label{nl12}
-\Gamma_p^0 \left[F(1,1)-F(1,0)\right] + \kappa_p F^{(1,0)}(1,1)=0.
\end{equation}
Using the fact that $F(1,1) = 1$ from the normalization of
$P^s_{r,p}$, and $F^{(1,0)}(1,1) = \langle r \rangle$, $F(1,0) =
\sum_{r=0}^\infty P^s_{r,p_0}$ from the definition of the generating
function in Eq.~\eqref{nl9}, we can use Eqs.~\eqref{nl11} and
\eqref{nl12} to find:
\begin{equation}\label{nl13}
\langle r \rangle = \frac{\Gamma_p^0 \xi}{\gamma_r + \kappa_p \xi}, \qquad 
\sum_{r=0}^\infty P^s_{r,p_0} = \frac{\gamma_r}{\gamma_r + \kappa_p \xi}.
\end{equation}
Thus we have an analytical expression for $\langle r \rangle$, one of
the moments necessary for calculating the Fano factor.  If we proceed
to the next order of derivation, applying $\partial_{z_1}^2$,
$\partial_{z_2}^2$, and $\partial_{z_1}\partial_{z_2}$ on
Eq.~\eqref{nl10} and evaluating at $z_1=1$, $z_2=1$, we can extract
from these three equations the following moment relations:
\begin{equation}\label{nl13b}
\begin{split}
&\langle p -p_0 \rangle = \frac{\kappa_p ((\gamma_r+\kappa_p)\xi-\Delta (\gamma_r+ \kappa_p \xi))}{\gamma_r(\gamma_r+\kappa_p \xi)},\\
&\sigma_r^2 = (1-\Delta)\langle r \rangle, \qquad \langle r (p-p_0) \rangle = (1-\Delta)\frac{\Gamma_p^0}{\gamma_r} - \frac{\langle r \rangle}{\xi},
\end{split}
\end{equation}
where $\Delta$ is defined as
\begin{equation}\label{nl13c}
\Delta = \langle r \rangle - \frac{\gamma_r + \kappa_p \xi}{\gamma_r}\sum_{r=0}^\infty r P^s_{r,p_0}.
\end{equation}
Thus the Fano factor $\sigma_r^2/\langle r \rangle = 1-\Delta$, but
unfortunately we do not have an explicit solution for $\Delta$ from
the generating function approach.  (Higher order partial derivatives
of Eq.~\eqref{nl10} do not form a closed system of equations.)
However, the moment relations in Eq.~\eqref{nl13b} will prove useful
below.

From Eq.~\eqref{nl13} we note that $\langle r \rangle \to \infty$ as
$\Gamma_p^0 \to \infty$, so the distribution is pushed toward larger
$r$ as the step functions become steeper, just as we saw in the
numerical optimization (Fig.~\ref{nlfig1}).  In the large $r$ limit,
we can approximate $P^s_{r,p}$ as a continuous function of $r$ (though
it remains discrete in $p$).  Based on the numerical optimization
results, we choose the following Gaussian ansatz for $P^s_{r,p_0}$,
the first non-negligible $p$ slice of the distribution:
\begin{equation}\label{nl14}
P^s_{r,p_0} = A_0 e^{-(r-\lambda_0)^2/(2 s_0^2)}.
\end{equation}
The parameters $\lambda_0$ and $s_0$ are to be determined, while $A_0$
must be chosen to satisfy $\sum_{r=0}^\infty P^s_{r,p_0}$ from
Eq.~\eqref{nl13}.  In the continuum, large $r$ limit we can
approximate the sum as $\sum_{r=0}^\infty P^s_{r,p_0} \approx
\int_{-\infty}^\infty dr\,P^s_{r,p_0}$, which implies that
\begin{equation}\label{nl15}
A_0 = \frac{\gamma_r}{\sqrt{2\pi s_0^2}(\gamma_r + \kappa_p \xi)}.
\end{equation}
Similarly, Eq.~\eqref{nl13c} gives
\begin{equation}\label{nl15b}
\Delta = \langle r \rangle - \lambda_0,
\end{equation}
so finding $\lambda_0$ is equivalent to finding $\Delta$.

Let us now show that the ansatz of Eq.~\eqref{nl14} yields a solution
$P^s_{r,p}$ for $p\ge p_0$ that satisfies Eq.~\eqref{nl7} in the large
$\Gamma_p^0$ limit.  Using Eq.~\eqref{nl8} and the continuum
approximation along the $r$ direction, we can rewrite Eq.~\eqref{nl7}
for $p\ge p_0$ as
\begin{equation}\label{nl16}
\begin{split}
&0 = {\cal R}_{r,p} \approx \gamma_r \partial_r (r P^s_{r,p}) - K_r^0
\delta_{p,p_0} \partial_r P^s_{r,p} + \Gamma_p^0 P^s_{r,p+1}\\
&- (1-\delta_{p,p_0})\Gamma_p^0 P^s_{r,p}+ \kappa_p r
\left[(1-\delta_{p,p_0})P^s_{r,p-1} - P^s_{r,p}\right].
\end{split} 
\end{equation}
Plugging the ansatz for $P^s_{r,p_0}$ from Eq.~\eqref{nl14} into
Eq.~\eqref{nl16} for $p=p_0$, we can solve for $P^s_{r,p_0+1}$,
\begin{equation}\label{nl17}\begin{split}
&P^s_{r,p_0+1} = \\
&A_0 e^{-(r-\lambda_0)^2/(2 s_0^2)}
 \frac{(\gamma_r r-K_r^0)(r-\lambda_0) + (\kappa_p r-\gamma_r)s_0}{\Gamma_p^0 s_0}.
\end{split}
\end{equation}
Similarly, once $P^s_{r,p_0}$ and $P^s_{r,p_0+1}$ are known,
Eq.~\eqref{nl16} for $p = p_0+1$ yields $P^s_{r,p_0+2}$,
\begin{equation}\label{nl18}
\begin{split}
&P^s_{r,p_0+2} = \frac{A_0 e^{-(r-\lambda_0)^2/(2 s_0^2)}}{{({\Gamma_p^0})^2 s_0^2}}\cdot\\
&\qquad\Bigl[ s_0^2 \left(-{\Gamma_p^0} \gamma_r+\gamma_r^2-3 \gamma_r \kappa_p r+\kappa_p^2 r^2\right)\\
&\qquad +s_0 \bigl\{{\Gamma_p^0} (\lambda_0-r) ({K_r^0}-\gamma_r r)\\
&\qquad +\lambda_0 \left(3 \gamma_r^2 r-\gamma_r \left(2 \kappa_p r^2+{K_r^0}\right)+\kappa_p {K_r^0} r\right)\\
&\qquad +r (2 \gamma_r-\kappa_p r) ({K_r^0}-2 \gamma_r r)\bigr\}\\
&\qquad+\gamma_r r (\lambda_0-r)^2 (\gamma_r r-{K_r^0})\Bigr].
\end{split}
\end{equation}
We can iterate this procedure, using Eq.~\eqref{nl16} to generate
analytical expressions for all $P^s_{r,p_0+m}$, $m>0$, which depend on
the unknown parameters $\lambda_0$ and $s_0$.  To solve for these
parameters, let us first enforce the normalization condition,
\begin{equation}\label{nl19}
1 = \sum_{m=0}^\infty \sum_{r=0}^\infty P^s_{r,p_0 + m} \approx \sum_{m=0}^\infty \int_{-\infty}^{\infty} dr\,P^s_{r,p_0+m}.
\end{equation}
Though tedious, the integrals on the right-hand side of
Eq.~\eqref{nl19} can be explicitly carried out for each $m$, since
$P^s_{r,p_0+m}$ has the form of a Gaussian
$\exp(-(r-\lambda_0)^2/(2s_0^2))$ times a polynomial in $r$.  Since we
are interested in the large $\Gamma_p^0$ limit, we can Taylor expand
the integrals up to first order in the small variable
$(\Gamma_p^0)^{-1}$, which gives the following result:
\begin{equation}\label{nl20}
\begin{split}
&\int_{-\infty}^{\infty} dr\,P^s_{r,p_0+m} \approx \frac{\gamma_r}{\gamma_r+\kappa_p \xi} \left(\frac{\kappa_p \xi}{\gamma_r+\kappa_p \xi} \right)^m+\\
& \frac{\gamma_r m (\kappa_p \xi)^{m} (\xi (-2 \Delta-\xi m+\xi)+(m-1) \tilde{s}_0 (\gamma_r+\kappa_p\xi))}{2 \Gamma_p^0 \xi^2  (\gamma_r+ \kappa_p\xi)^{m}},
\end{split}
\end{equation}
where $\tilde{s}_0 = s_0 /\Gamma_p^0$, and we have used
Eq.~\eqref{nl15b} to write $\lambda_0 = \langle r \rangle -\Delta$,
and Eq.~\eqref{nl13} for $\langle r \rangle$.  Plugging
Eq.~\eqref{nl20} into Eq.~\eqref{nl19} and carrying out the sum over
$m$, the normalization condition becomes
\begin{equation}\label{nl22}
1 = 1 - \frac{\kappa_p (\gamma_r+ \kappa_p\xi) \left(\Delta \gamma_r-\kappa_p s_0 (\gamma_r+\kappa_p\xi)+\kappa_p\xi^2\right)}{\Gamma_p^0 \gamma_r^2}.
\end{equation}
Thus the term of order $(\Gamma_p^0)^{-1}$ on the right must be zero, implying the following relation between $\tilde{s}_0$ and $\Delta$,
\begin{equation}\label{nl23}
\tilde{s}_0 = \frac{\Delta \gamma_r + \kappa_p \xi^2}{\kappa_p(\gamma_r + \kappa_p \xi)}.
\end{equation}
In order to complete the derivation and solve for $\Delta$, we need
to calculate the moment $\langle p - p_0 \rangle$,
\begin{equation}\label{nl24}
\langle p - p_0 \rangle = \sum_{m=0}^\infty \sum_{r=0}^\infty m P^s_{r,p_0 + m} \approx \sum_{m=0}^\infty m \int_{-\infty}^{\infty} dr\,P^s_{r,p_0+m}.
\end{equation}
Plugging in Eq.~\eqref{nl20} for the integral, we carry out the sum
over $m$ and simplify using Eq.~\eqref{nl23}, giving
\begin{equation}\label{nl24}
\langle p - p_0 \rangle = \frac{\kappa_p (\Gamma_p^0 \gamma_r \xi + \Delta (\gamma_r + \kappa_p \xi)^2)}{\Gamma_p^0 \gamma_r^2}.
\end{equation}
Setting this equal to the $\langle p -p_0 \rangle$ result from
Eq.~\eqref{nl13b}, we finally can solve for $\Delta$, or equivalently the Fano factor $\sigma_r^2/\langle r \rangle = 1-\Delta$,
\begin{equation}\label{nl25}
\begin{split}
\frac{\sigma_r^2}{\langle r \rangle} &= 1- \frac{\Gamma_p^0 \gamma_r \kappa_p (1-\xi)\xi}{(\gamma_r+ \kappa_p \xi)(\Gamma_p^0 \gamma_r + (\gamma_r+\kappa_p \xi)^2)}\\
& \approx 1-\frac{(1-\xi)\xi \kappa_p}{\gamma_r+\kappa_p \xi} + {\cal O}((\Gamma^0_p)^{-1}),
\end{split}
\end{equation}
keeping the leading terms in the Taylor expansion for small
$(\Gamma^0_p)^{-1}$.  The Fano factor achieves a minimum value equal
to the WK linear optimum,
\begin{equation}\label{nl26}
\frac{\sigma_{r,\text{min}}^2}{\langle r \rangle} = \frac{2}{1+\sqrt{1+B}} = \frac{\sigma_{r,\text{WK}}^2}{\langle r \rangle}
\end{equation}
 at $\xi = \xi_\text{min} = 1/(1+\sqrt{1+B})$, where $B=\kappa_p /
 \gamma_r$.  Thus we see explicitly that nonlinear threshold
 regulation with $K_r(p)$ and $\Gamma_p(p)$ behaving like step
 functions can directly match (but not improve on) the efficiency of
 the optimal WK linear filter, so long as $\Gamma_p^0$ is large and
 the ratio of the step function heights assumes a particular value
 $\xi_\text{min}$.  Counterintuitively, this occurs despite the fact
 that the $p$ copy numbers can be very small in our system, with a
 narrow range of fluctuations in which discreteness plays a major
 role.

\section{Optimality for the TetR gene network under extrinsic noise}

In the frequency domain, we will model $n^\text{ext}_\alpha(\omega)$,
the extrinsic part of the noise associated with species $\alpha$
using,
\begin{equation}\label{eq14}
n^\text{ext}_\alpha(\omega) = \frac{\sqrt{2 c_\alpha \bar{k}_\alpha}}{1-i\omega \tau_e} \eta^\text{ext}_\alpha(\omega), 
\end{equation}
where $c_\alpha$ is a coefficient measuring the strength of the noise,
and $\eta^\text{ext}(\omega)$ is a Fourier-space Gaussian white noise
function.  Comparing to the definition of the intrinsic noise,
$n^\text{int}_\alpha(\omega) = \sqrt{2
  \bar{k}_\alpha}\eta_\alpha(\omega)$, we see that $c_\alpha$ is the
ratio of the extrinsic to intrinsic noise PSD for species $\alpha$ at
$\omega = 0$.  The $(1-i\omega \tau_e)^{-1}$ factor acts as a cutoff
that suppresses frequencies $\omega \gg \tau_e^{-1}$.  The total noise
function for species $\alpha$ is the sum of intrinsic and extrinsic
contributions, $n_\alpha(\omega) = n^\text{int}_\alpha(\omega) +
n^\text{ext}_\alpha(\omega)$.  We will focus on how the addition of
extrinsic noise affects the optimality conditions using the TetR yeast
gene circuit example.

The calculation of $H_{\rm opt}(\omega)$ proceeds analogously to the
no-extrinsic-noise procedure described in the
main text. The power spectra of the signal and noise are,
\begin{equation}\label{si8}
\begin{split}
P_s(\omega) &= 2 \bar{r} \tau_r \left[\frac{1}{1 + (\omega \tau_r)^2} + \frac{c_r}{(1 + (\omega \tau_r)^2)(1+(\omega \tau_e)^2)}\right], \\
P_n(\omega) &= \frac{2\bar{r}\tau_r}{B}\left[1 + \frac{c_p}{1+(\omega \tau_e)^2}\right].
\end{split}
\end{equation}
The first and second terms in the square brackets represent the
intrinsic and extrinsic contributions respectively. The latter is
parameterized by the coefficients $c_r$ and $c_p$, and the timescale
$\tau_e$, which is assumed to be much larger than the dominant
timescale, $\tau_r$, characterizing the $R$ fluctuations.  The signal
plus noise power spectrum, $P_y(\omega) = P_s(\omega) + P_n(\omega)$,
can be rewritten as a causal decomposition in the following manner:
\begin{equation}\label{si9}
\begin{split}
P_y(\omega) &= \left|\left(\frac{2\bar{r}\tau_r}{B}\right)^{1/2} \frac{(\rho_+ - i \omega \tau_r)(\epsilon^{-1} \rho_- - i \omega \tau_e)}{(1- i\omega \tau_r)(1-i\omega \tau_e)}\right|^2\\
&\equiv |P_y^c(\omega)|^2,
\end{split}
\end{equation}
where $\epsilon \equiv \tau_r/\tau_e$, and
\begin{equation}\label{si10}
\begin{split}
\rho_{\pm} &= \sqrt{\frac{\mu\pm \sqrt{\mu^2-4\epsilon^2 \nu}}{2}},\\
\mu &= 1+B + \epsilon^2(1+c_p),\\
\nu &= 1+B(1+c_r) + c_p. 
\end{split}
\end{equation}
The expression $P_s(\omega)/P_y^c(\omega)^\ast$ and its additive
causal decomposition $\{P_s(\omega)/P_y^c(\omega)^\ast\}_c$ is given
by:
\begin{equation}\label{si11}
\begin{split}
&\frac{P_s(\omega)}{P_y^c(\omega)^\ast} =\\
&\qquad \frac{(2\bar{r} \tau_r B)^{1/2}(1+ c_r + (\omega \tau_e)^2)}{(1-i \omega \tau_r)(1-i \omega \tau_e) (\rho_+ + i \omega \tau_r) (\epsilon^{-1}\rho_- + i \omega \tau_e)},
\end{split}
\end{equation}
\begin{equation}\label{si12}
\begin{split}
&\left\{\frac{P_s(\omega)}{P_y^c(\omega)^\ast}\right\}_c =\\
&\qquad \frac{(2\bar{r} \tau_r B)^{1/2}(1+ c_r - \epsilon^{-2})}{(1-i \omega \tau_r)(1- \epsilon^{-1}) (\rho_+ + 1) (\epsilon^{-1}\rho_- + \epsilon^{-1})}\\
&\qquad + \frac{(2\bar{r} \tau_r B)^{1/2}c_r}{(1- \epsilon)(1-i \omega \tau_e) (\rho_+ + \epsilon) (\epsilon^{-1}\rho_- + 1)}.
\end{split}
\end{equation}
Using Eqs.~\eqref{si12} and \eqref{si9} in Eq.~\eqref{eq7}, we 
obtain the form for the optimal filter function:
\begin{equation}\label{si13}
H_{\rm opt}(\omega) =  \frac{B K(\omega)}{(1-\epsilon) (\rho_+ - i \omega \tau_r) (\epsilon^{-1}\rho_- - i \omega \tau_e)},
\end{equation}
where
\begin{equation}\label{si14}
\begin{split}
K(\omega) &= \frac{1-(1+c_r)\epsilon^2}{(1+\rho_-)(1+\rho_+)}(1-i \omega \tau_e)\\& + \frac{c_r \epsilon}{(\epsilon+\rho_-)(\epsilon+\rho_+)}(1-i\omega \tau_r).
\end{split}
\end{equation}
Since $\epsilon$ is presumed small, we will expand $H_{\rm opt}$ to
lowest order in $\epsilon$, giving the approximate expression:
\begin{equation}\label{si15}
H_{\rm opt}(\omega) \approx \frac{\sqrt{1+B}-1}{\sqrt{1+B} - i \omega \tau_r}\cdot \frac{1+c_r \frac{1+\sqrt{1+B}}{\sqrt{\nu}+\sqrt{1+B}}-i \omega \tau_e}{\sqrt{\frac{\nu}{1+B}}-i \omega \tau_e}.
\end{equation}
The first rational term is just the optimal filter result in the
intrinsic-only case, Eq.~\eqref{si6}, while the second term represents
the modification needed to accommodate the extrinsic noise.  As
expected, the latter term approaches $1$ when $c_r, c_p \to 0$, since
$\nu \to 1+B$ in this limit.

There is a different non-trivial scenario where the second term is equal to 1. If the
noise magnitudes $c_r$ and $c_p$ are related such that,
\begin{equation}\label{si16}
1+c_r \frac{1+\sqrt{1+B}}{\sqrt{\nu}+\sqrt{1+B}} = \sqrt{\frac{\nu}{1+B}},
\end{equation}
then the numerator and denominator exactly cancel each other out,
removing the $\tau_e$ dependence from the optimal filter.  Using the
definition $\nu = 1+B(1+c_r)+c_p$, Eq.~\eqref{si16} can be simplified to
yield the relation:
\begin{equation}\label{si17}
c_r = \frac{1}{1+\sqrt{1+B}} c_p.
\end{equation}
If this condition is satisfied, $H_{\rm opt}(\omega)$ is identical to
the intrinsic-only optimal filter of Eq.~\eqref{si6} (to lowest order in
$\epsilon$), and hence the approximate optimality is also achieved at
the same feedback value, $G^{\rm opt}_{rp} \approx {\cal G}^{\rm
  opt}_{rp} (B,\tau_p)$.

Thus, the yeast gene circuit can still be fine-tuned to approach a WK
optimal filter even in the presence of extrinsic noise.  However, this
tuning requires the relative strengths $c_r$ and $c_p$ of the R and P
extrinsic noise to be related (at least approximately) by Eq.~\eqref{si17}.  The resulting
minimal possible Fano factor $\sigma^2_{r,\text{opt}}/\bar{r}$ is:
\begin{equation}\label{eq17}
\frac{\sigma^2_{r,\text{opt}}}{\bar{r}} \approx \frac{2}{1+\sqrt{1+B}} + \frac{\tau_r}{\tau_e(1+B+\sqrt{1+B})}c_p.
\end{equation}
This is the intrinsic-only result of Eq.~\eqref{eq11} in the main text
plus an extrinsic noise contribution in the second term. Not
surprisingly, with more total noise in the system, the standard
deviation of the optimally filtered output increases.  Since the
second term is of the order $\tau_r/\tau_e$ it follows that the bigger
the difference in time scales between the extrinsic noise ($\tau_e$)
and the mRNA dynamics ($\tau_r$), the easier it is to filter out the
extrinsic influence on the mRNA fluctuations.  For $B \gg 1$, the
fundamental limit on the noise suppression still arises from the
intrinsic term in $\sigma^2_{r,\text{opt}}/\bar{r}$, which scales like
$\sim B^{-1/2}$; the extrinsic contribution decays more rapidly,
$\sim B^{-1}$.

The blue curves in Fig.~\ref{extr} show the linear theory predictions
for $\sigma^2_r/\bar{r}$ as a function of $A$ in two cases: (i) $c_p =
80$, $c_r = 23$; (ii) $c_p = 160$, $c_r = 46$.  The burst ratio $B=5$,
and $\tau_e$ is set equal to $\gamma_p^{-1}$, the longest time scale
among the experimentally fitted parameters.  For both these cases
the noise strengths $c_p$ and $c_r$ satisfy the relation in
Eq.~\eqref{si17}, and hence it is possible to tune the system to
approximately achieve WK optimality, just as in the intrinsic-only
scenario.  The noise magnitudes were chosen so that the system is
noticeably perturbed by the extrinsic contribution. For example, if
the signal $s(t)$ is split into intrinsic and extrinsic parts
$s^\text{int}(t)$ and $s^\text{ext}(t)$, the ratios of their
respective standard deviations are
$\sigma_s^\text{ext}/\sigma_s^\text{int} = 0.8$ for case (i) and $1.6$
for case (ii).  The value of $\sigma^2_{r,\text{opt}}/\bar{r}$ is
marked by horizontal dashed lines, and the point $A = A_\text{opt}$,
where $G^\text{opt}_{rp}(\omega) \approx {\cal
  G}^\text{opt}_{rp}(B,\tau_p)$ is satisfied, by a filled circle.  In
all cases the system approaches $\sigma^2_{r,\text{opt}}/\bar{r}$ near
$A = A_\text{opt}$, verifying the optimality prediction.

As in the intrinsic-only scenario discussed in the main text, we can
test the usefulness of the linear theory through Gillespie simulations
(results shown as open squares and circles in Fig.~\ref{extr}), and reach a
similar conclusion even in the presence of extrinsic noise. At large
volumes, $V = 10V_0$, the simulations converge to the linear theory,
whereas for the more realistic volume $V=V_0$ we see discrepancies due
to nonlinearity and low copy numbers ($V_0 = 60$ fL).  Nevertheless,
the Fano factor still reaches a minimum close to the predicted
$A_\text{opt}$ and $\sigma^2_{r,\text{opt}}/\bar{r}$ values.

\end{document}